\documentclass[letterpaper,english,aps,prb,amsfonts,amssymb,amsmath,longbibliography,superscriptaddress]{revtex4-2}
\usepackage{balance}
\usepackage[T1]{fontenc}
\usepackage[utf8]{inputenc}
\usepackage[english]{babel}
\usepackage{graphics}
\usepackage{amssymb}
\usepackage{amsmath, esint}
\usepackage{overpic}
\usepackage[toc]{appendix}
\usepackage{bbold}

\usepackage{natbib}
\usepackage{xcolor}
\usepackage{soul}
\usepackage{commath}
\usepackage{tikz}
\usepackage{physics}

\usepackage{hyperref}

\newcommand{\be}{\begin{equation}}
	\newcommand{\ee}{\end{equation}}
\newcommand{\bspl}{\begin{split}}
	\newcommand{\espl}{\end{split}}
\newcommand{\bea}{\begin{eqnarray}}
	\newcommand{\eea}{\end{eqnarray}}

\def\pd{\partial}

\def\bk{{\bf k}}

\def\nn{\nonumber}
\def\lb{\label}
\def\pref#1{(\ref{#1})}

\newcount\bozza \bozza=0
\ifnum\bozza=1
\newdimen\shift \shift=-1.5truecm
\def\lb#1{%
	{\label{#1}\rlap{\kern\shift{$\scriptstyle#1$}}}}
\else\def\lb#1{\label{#1}} \fi

\newcommand\beq{\begin{equation}}
	\newcommand\eeq{\end{equation}}

\begin{document}
	
	\title{Optical absorption in tilted geometries as an indirect measurement of longitudinal plasma waves in layered cuprates}
	
	\author{Niccolò Sellati}
	\email{niccolo.sellati@uniroma1.it}
	\affiliation{Department of Physics and ISC-CNR, ``Sapienza'' University of Rome, P.le
		A. Moro 5, 00185 Rome, Italy}
	\author{Jacopo Fiore}
	\affiliation{Department of Physics and ISC-CNR, ``Sapienza'' University of Rome, P.le
		A. Moro 5, 00185 Rome, Italy}
	\author{Claudio Castellani}
	\affiliation{Department of Physics and ISC-CNR, ``Sapienza'' University of Rome, P.le
		A. Moro 5, 00185 Rome, Italy}
	\author{Lara Benfatto}
 	\email{lara.benfatto@roma1.infn.it}
	\affiliation{Department of Physics and ISC-CNR, ``Sapienza'' University of Rome, P.le
		A. Moro 5, 00185 Rome, Italy}

\begin{abstract}

Electromagnetic waves  propagating in a layered superconductor with arbitrary momentum with respect to the main crystallographic directions display an unavoidable mixing between longitudinal and transverse degrees of freedom. Here we show that this basic physical mechanism explains the emergence of a well-defined absorption peak in the in-plane optical conductivity for light propagating at small tilting angles with respect to the stacking direction in layered cuprates. More specifically, we show that this peak, often interpreted as a spurious leakage of the $c$-axis Josephson plasmon, is instead a signature of the true longitudinal plasma mode occurring at larger momenta. By combining a classical approach based on Maxwell's equations with a full quantum derivation of the plasma modes based on the modelling of the superconducting phase degrees of freedom, we provide an analytical expression for the absorption peak as a function of the tilting angle and light polarization. We suggest that an all-optical measurement in tilted geometry can be used as an alternative way to access  plasma-wave dispersion, usually measured by means of large-momenta scattering techniques like resonant inelastic X-ray scattering (RIXS) or electron energy loss spectroscopy (EELS). 

\end{abstract}
\date\today
\maketitle
	
\section{Introduction}
In superconductors the breaking of the continuous gauge symmetry below the superconducting (SC) critical temperature is accompanied by the emergence of two collective modes, associated with the amplitude (Higgs) or phase (Goldstone) fluctuations of the complex SC order parameter, whose absolute value at equilibrium defines the spectral gap for single-particle excitations \cite{nagaosa}. While the former is a massive excitation, the latter is massless at long wavelength, reflecting the infinity of possible ground states connected by a global change of the order-parameter phase. Nonetheless, the coupling of the SC phase to the electron density is directly affected by long-range Coulomb interactions between charged electrons. This effect moves the phase mode to the plasma energy scale \cite{anderson_pr58}, that is usually much larger than the spectral gap. As a consequence, optical signatures at the plasma energy scale, i.e.\ at the zero of the dielectric function, are usually unaffected by the SC transition. A rather different phenomenology is instead observed in anisotropic layered superconductors, i.e.\ systems where the pairing mainly occurs within planes stacked along the $c$ direction, and SC order is established below $T_c$ thanks to a weak Josephson-like inter-plane interaction. The hallmark of this category is represented by high-temperature cuprates \cite{keimer_review15}, where the marked anisotropy has been experimentally proven by different optical probes, starting from linear optics, which measures two well-separated energy scales for the plasma modes at long-wavelength for electric fields propagating in the CuO$_2$ planes or perpendicular to them. In these systems the incoherent quasiparticle hopping along the stacking direction makes the $c$-axis response badly metallic: in contrast, below $T_c$ the opening of a sizeable spectral gap along with the weak inter-layer pair hopping leave a rather sharp SC plasma edge at a frequency $\omega_c$ of few THz in the optical reflectivity, that clearly testifies the emergence of a well-defined SC Josephson plasmon. Even though this feature has been experimentally observed already in the late 1990s \cite{homes_prl93,vandermarel96,yamada_prl98,erb_prl00,vandermarel_prb01,bernhard_prl11},
renewed interest in the physics of Josephson plasmons emerged more recently. Such interest has been triggered both by the applications to nano-plasmonic \cite{basov_prb14,gozar_nqm21} and by the role played by Josephson plasmons in non-linear THz spectroscopy \cite{nori_review10,cavalleri_nmat14,cavalleri_science18,wangNL_prx20,shimano_prb23,averitt_prb23}. In both cases it becomes theoretically relevant understanding the momentum dependence of the plasmon dispersion at generic momentum, i.e.\ not along the main crystallographic axes. In this configuration one immediately realizes that the anisotropy leads to a non-trivial response of the system, due to the fact that the current induced by the external electric field is no more parallel to the field itself. As it has been extensively discussed in details in Refs.\ \cite{gabriele_prr22,sellati_prb23,gabriele_prb24}, this mechanism leads to a mixing of the longitudinal and transverse response inside the material, making the distinction between plasmons and polaritons blurred at momenta smaller than a scale $\bar k\sim \sqrt{\omega_{ab}^2-\omega_c^2}/c$ set by the anisotropy between in-plane $\omega_{ab}$ and out-of-plane $\omega_c$  plasma frequencies. Since usually $\omega_{ab}\gg \omega_c$, the effect is relevant for non-linear Josephson plasmonics in the THz regime \cite{bulaevskii_prb94,bulaevskii_prb02,cavalleri_review,demler_cm24}, but does not affect e.g.\ the measurements of plasmons in RIXS \cite{lee_rixs_nature18,liu_rixs_npjqm20,zhou_prl20,huang_rixs_prb22,hepting_rixs_cm23} or EELS \cite{mitrano_pnas18,mitrano_prx19,abbamonte_natcomm23}, that usually measure momenta in a fraction of the Brillouin-zone. In the present manuscript we investigate an additional consequence of the above-mentioned mixing, showing how even linear optics can be used to disentangle the longitudinal-transverse mixing in a reflection or transmission geometry which highlights the emergence inside the material of a longitudinal response induced by an external transverse electromagnetic wave. The effect manifests as an absorption peak at a scale nearby $\omega_c$ for an electromagnetic wave travelling at small angle with respect to the $c$ direction. This feature has been measured in the past in different samples of electron-doped cuprates \cite{pimenov_apl00,pimenov_prb00,pimenov_prb02,armitage_prb21} below $T_c$, and it has been often interpreted as a leakage of the $c$-axis plasmon into the in-plane response \cite{vandermarel_comment93}. Even more interestingly, the peak position has been shown to change by varying the wave polarization in the plane of incidence, challenging considerably the interpretation of the results. Here we provide a full theoretical description of the microscopic mechanism behind the anomalous absorption peak, and we show that it is a direct consequence of the plasmon-polariton mixing in an anisotropic layered superconductor. We argue that this effect can be used to indirectly probe the plasmon dispersion that usually appears in RIXS and EELS experiments at much larger momenta and, by changing the light polarization, to extract the in-plane and out-of-plane plasma frequencies. Our findings are benchmarked against existing experimental data for cuprates. On a more general ground, our results offer a novel perspective on the possibility to access collective polariton modes in complex materials by properly engineering optical measurements. 

\section{Anisotropic linear response of layered system}
\begin{figure}[t!]
\centering
\includegraphics[width=10.5cm]{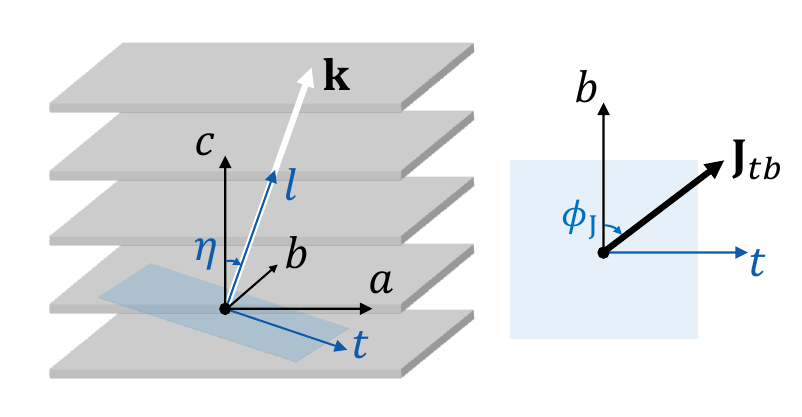}
\caption{\label{fig1} Sketch of the notation used in the manuscript to define the reference frames. The crystalline orientation defines the frame $(a,b,c)$, the direction of the momentum defines $(t,b,l)$. The angles $\eta$ and $\phi_\text J$ are also represented. The $tb$-plane is highlighted in blue. }
\end{figure}

As we discussed in the Introduction, several experiments in electron-doped cuprates \cite{pimenov_apl00,pimenov_prb00,pimenov_prb02,armitage_prb21} have shown the emergence of a peak in the in-plane conductivity below $T_c$ at a frequency close to the one of the out-of-plane plasma edge, whose position moves by changing the light polarization. This peak is often interpreted as a spurious effect due to the leakage of the $c$-axis plasmon into the in-plane response \cite{vandermarel_comment93}, and light polarization is used to remove the effect \cite{armitage_prb21}. However, in Ref.\ \cite{pimenov_prb02} the problem has been investigated in details by growing on purpose a sample with a stacking direction tilted with respect to the light wave-vector, and a preliminary interpretation has been provided linked to such a tilted geometry. Here we will follow the same reasoning, and we will study the response for propagating wave-vector at tilted angle with respect to the stacking direction. To fix the notation, in the following we will use the convention by which the SC sheets are parallel to the $ab$-plane and stacked along the $c$-axis. We then assume, without loss of generality, that the momentum $\textbf k$ of the propagating wave is along the $ac$-plane ($k_b=0$). The angle between $\textbf k$ and the $c$-axis is denoted $\eta$ and the angle between the transverse current and the $b$-axis is denoted $\phi_\text J$ (see Fig.\ \ref{fig1} for the notation followed in this manuscript). Even though in such tilted geometries the discussion of the Fresnel conditions at the sample/air boundary is not straightforward, we will postpone this analysis to the last Section, and we will focus here on the behavior inside the sample. We are then interested in determining the measured conductivity, defined as the ratio between the current $\textbf J$ induced in the field direction 
and the modulus of the electric field $\textbf E$ itself.\\
Because of anisotropy, the charge mobility within the planes is much higher than in between stacked layers and the current $\textbf J$ in the material is in general not parallel to $\textbf E$ unless propagation occurs along the the principal axes of the crystal $(a,b,c)$. Indeed, in general one can write the conductivity tensor as 
\begin{align}\lb{jabc}
    \begin{pmatrix}\text J_{a} \\ \text J_{b} \\ \text J_c\end{pmatrix}=
    \begin{pmatrix}\sigma_{ab} & 0 & 0 \\ 
    0 & \sigma_{ab} & 0 \\
    0 & 0 & \sigma_c\end{pmatrix}
    \begin{pmatrix}\text E_{a} \\ \text E_b \\ \text E_c \end{pmatrix},
\end{align}
where $\sigma_{ab}$ and $\sigma_c$ are the in-plane and out-of-plane conductivities respectively.  In the following we simplify the tensorial notation by writing the reference frame in which a quantity is considered as its subscript, e.g.\ Eq. \eqref{jabc} reads $\textbf{J}_{abc}=\hat{\boldsymbol{\sigma}}_{abc}\textbf{E}_{abc}$. If the wave is propagating perpendicularly to the planes ($\eta=0$), the electric field oscillates within the SC sheets and one directly extracts $\sigma_{ab}$ from the measured transmissivity/reflectivity; analogously, with a wave propagating within the planes ($\eta=\pi/2$) one can measure $\sigma_c$. However, for a generic value of the propagation angle the measured conductivity will be a combination of the two quantities. In other words, for $\textbf k$ at generic angle $\eta$  the current $\textbf J$ will develop both longitudinal and transverse components with respect to the momentum. 
%
To see this explicitly, we perform a
rotation of angle $\eta$ around the $b$-axis to move in the reference frame $(t,b,l)$ in which $l$ labels the longitudinal components and $t$ labels the transverse component with respect to the momentum in the $ac$-plane, while preserving the second transverse component $b$. In this frame Eq.\ \eqref{jabc} transforms into $\textbf{J}_{tbl}=\hat{\boldsymbol{\sigma}}_{tbl}\textbf{E}_{tbl}$, where the conductivity tensor now reads
\begin{align}\lb{jlbt}
    \hat{\boldsymbol{\sigma}}_{tbl}=
    \begin{pmatrix}\sigma_{ab}\cos^2\eta+\sigma_c\sin^2\eta & 0 & (\sigma_c-\sigma_{ab})\sin\eta\cos\eta \\ 
    0 & \sigma_{ab} & 0 \\
    (\sigma_c-\sigma_{ab})\sin\eta\cos\eta & 0 & \sigma_{ab}\sin^2\eta+\sigma_c\cos^2\eta\end{pmatrix}.
\end{align}
Notice that the components $\text J_t$ and $\text J_l$ are coupled to both $\text E_t$ and $\text E_l$, as one expects in an anisotropic crystal, whereas the transverse $\text J_b$ component only couples to $\text E_b$. As we will detail below, what one determines experimentally is an effective conductivity defined as the ratio between the transverse current and the transverse electric field. According to Ampere's law $\frac{4\pi}{c}\textbf J+\frac{\varepsilon_\infty}{c}\frac{\partial\textbf E}{\partial t}=0$, with $c$ the light velocity and $\varepsilon_\infty$ the background dielectric constant, so that the current in the longitudinal direction is compensated by the displacement current. We then obtain the relation $4\pi \text J_l-i\omega\varepsilon_\infty\text E_l=0$ that can be used to eliminate the longitudinal component and write a system that only takes the transverse components $t$ and $b$ into consideration, $\textbf J_{tb}=\hat{\boldsymbol{\sigma}}_{tb}\textbf E_{tb}$. The transverse conductivity tensor reads
%
\begin{align}\lb{jbt}
    \hat{\boldsymbol{\sigma}}_{tb}=\begin{pmatrix}
        \sigma_{t} & 0 \\
        0 & \sigma_{ab}
    \end{pmatrix},
\end{align}
where 
\begin{align}\lb{sigmat}
    \sigma_t(\omega,\eta)=\frac{-\frac{i\omega\varepsilon_\infty}{4\pi}\big(\sigma_{ab}\cos^2{\eta}+\sigma_c\sin^2\eta\big)+\sigma_{ab}\sigma_{c}}{-\frac{i\omega\varepsilon_\infty}{4\pi}+\sigma_{ab}\sin^2\eta+\sigma_c\cos^2\eta}.
\end{align}
For an electric field polarized along $t$ ($\text E_b=0$), Eq.\ \pref{sigmat} immediately gives the conductivity we were looking for, $\sigma_t=\text J_t/\text E_t$. This expression was first derived in Ref.\ \cite{pimenov_prb00}, and its real part displays a peak with central frequency that moves with $\eta$. To show it explicitly, we replace $\sigma_{ab}=\frac{i\omega}{4\pi }\frac{\varepsilon_\infty\omega_{ab}^2}{(\omega+i0^+)^2}$ and $\sigma_{c}=\frac{i\omega}{4\pi }\frac{\varepsilon_\infty\omega_{c}^2}{(\omega+i0^+)^2}$, where $\omega_{ab}$ and $\omega_c$ are the in-plane and out-of-plane plasma frequencies respectively: one then immediately sees that the real part of $\sigma_t(\omega,\eta)$ peaks at frequency $\omega_l(\eta)$, that reads 
\begin{align}\lb{rpadisp}
    \omega_l^2(\eta)=\omega_{ab}^2\sin^2\eta+\omega_c^2\cos^2\eta.
\end{align}
As we will discuss below, $\omega_l$ does {\em not} define a plasma mode of the system: this can be immediately understood already within a classical approach, by writing explicitly the dielectric function corresponding to the conductivity \pref{sigmat}. By using $\sigma_{ab}=-\frac{i\omega}{4\pi}(\varepsilon_{ab}-\varepsilon_\infty)$ and $\sigma_{c}=-\frac{i\omega}{4\pi}(\varepsilon_{c}-\varepsilon_\infty)$ we can write the in-plane $\varepsilon_{ab}$ and out-of-plane $\varepsilon_c$ dielectric functions of the SC system as:
\begin{align}\lb{inpl1}
    \varepsilon_{ab}(\omega)=\varepsilon_\infty\left(1-\frac{\omega_{ab}^2}{(\omega+i0^+)^2}\right),
\end{align}
and
\begin{align}\lb{outpl1}
    \varepsilon_{c}(\omega)=\varepsilon_\infty\left(1-\frac{\omega_{c}^2}{(\omega+i0^+)^2}\right).
\end{align}
Thus Eq.\ \eqref{sigmat} can be recast as $\sigma_t=-\frac{i\omega}{4\pi}(\varepsilon_{t}-\varepsilon_\infty)$, where
\begin{align}\lb{epsit1}
    \varepsilon_t(\omega,\eta)=\varepsilon_\infty\frac{\big((\omega+i0^+)^2-\omega_{ab}^2\big)\big((\omega+i0^+)^2-\omega_c^2\big)}{(\omega+i0^+)^2\big(\omega^2-\omega^2_l(\eta)\big)}.
\end{align}
%
In Eq.\ \pref{epsit1} the frequency $\omega_l(\eta)$ in Eq.\ \eqref{rpadisp} appears as a divergence of the dielectric function, while the plasma frequencies in the long-wavelength limit appear as usual as zeros of the dielectric function. This already proves that the scale $\omega_l$ does not identify a true plasma mode. However, as we will demonstrate below, it turns out that  $\omega_l$ provides a good approximation for the {\em finite-momentum} longitudinal plasmon of the layered system {\em at large momenta}, i.e. in the momentum region where retardation effects are no more relevant.  As a consequence, the present results show that the optical absorptive peak in the tilted geometry, that appears as linear response in the long-wavelength limit,  can be used to indirectly access the plasma-wave dispersion at large momenta. Notice that in principle Eq.\ \eqref{sigmat} is valid in general for any collective mode in an anisotropic uniaxial system, provided that the corresponding expressions of $\sigma_{ab}(\omega)$ and $\sigma_c(\omega)$ are used.\\ 
Even though these considerations solve the problem of defining a transverse conductivity at tilted angles for electric field polarized along $t$, two main issues remain open. The first one regards the connection between the frequency of the peak \pref{rpadisp} and the real plasma modes of the anisotropic superconductor. The second point is to link these results to the measured quantity in an experiment with a generic polarization of the electric field. The first matter will be discussed in the next Section, by using a quantum formalism based on the description of the electromagnetic modes via the SC phase degree of freedom. The second issue will be the subject of the last Section, where we will explicitly study the Fresnel problem for transmission/reflection through a tilted-grown sample and we further discuss the dependence of the measurement on the polarization $\phi$ of the external incident electric field.

\section{Linear response of generalized plasma modes}
\subsection{Effective action description of plasma modes}
To gain a better physical insight into the results of the previous section, we will take advantage of the description of the plasma modes in the SC state obtained via the phase degrees of freedom. Indeed, as it has been recently discussed in Refs.\ \cite{gabriele_prr22,sellati_prb23}, this is a rather powerful and elegant approach to describe the interplay between longitudinal and transverse plasma waves in a layered superconductor, that leads to  generalized plasma modes with mixed character at low momenta. We summarize here the main ingredients of the derivation, referring the reader to Refs.\ \cite{gabriele_prr22,sellati_prb23} and references therein for a detailed derivation of the layered phase-only model. \\
Below the critical temperature $T_c$ the neighboring SC planes interact with a Josephson-like coupling \cite{nori_review10,cavalleri_review,keimer_review15,shibauchi_prl94,panagopoulos_prb96,bonn_prl04,fazio_review01} that is much weaker than the in-plane phase stiffness. Following the notation set above, we denote the in-plane superfluid stiffness by $D_{ab}$ and the out-of-plane one by $D_c$ and we write the Gaussian action for the phase fluctuations $\theta$ as \cite{randeria_prb00,benfatto_prb01,millis_prr20}
\begin{align}\lb{supfluid}
S_G[\theta]=\frac{1}{8}\sum_{q}\big[\kappa_0\Omega_m^2+D_{ab}k_{ab}^2+D_c k_c^2\big]|\theta(q)|^2,
\end{align}
where $q=(i\Omega_m,\textbf k)$ is the imaginary-time 4-momentum, with $\Omega_m=2\pi m T$ the bosonic Matsubara frequencies, $ k_{ab}=\sqrt{ k_a^2+ k_b^2}$ and $k_c$ are the in-plane and out-of-plane momentum respectively and $\kappa_0$ is the compressibility. In the following we denote by $\lvert\mathbf{k}\rvert^2=k_{ab}^2+k_{c}^2$.
We introduce the electromagnetic field $\textbf A$ by performing in Eq.\ \eqref{supfluid} the minimal coupling substitution $i\textbf k\theta\to i\textbf k\theta+2e\textbf A/c$, where $-e$ is the charge of the electron, and we also add the action of the free electromagnetic field \cite{nagaosa},
\begin{align}\lb{emfree}
S_{\text{e.m.}}[\textbf A]=\frac{1}{8\pi c^2}\sum_{q}\big[\varepsilon_\infty\Omega_m^2|\textbf A(q)|^2+c^2|\textbf k \times \textbf A(q)|^2\big],
\end{align}
Both the minimal coupling substitution and Eq.\ \eqref{emfree} are written in the Weyl gauge in which the scalar potential is zero. We then recast the coupling between the phase fluctuations and the electromagnetic field by performing the substitution
%
\begin{align}\lb{gauginv}
    \boldsymbol\psi(q)=i\textbf k\theta(q)+\frac{2e}{c}\textbf A(q).
\end{align}
These gauge-invariant fields provide a full description of the plasma modes once the phase fluctuations are integrated out \cite{gabriele_prr22,sellati_prb23}. To provide simple analytical expressions, in the following, we consider the limit for infinite compressibility. This is a good approximation in single-layer cuprates, as the effects of finite compressibility on the properties of the generalized plasma modes are negligible at small momenta \cite{gabriele_prr22}. Interestingly, these effects are actually crucial when studying the optical absorptive peak of bilayer superconductors \cite{homes_prl93,vandermarel96,yamada_prl98,erb_prl00,vandermarel_prb01,bernhard_prl11,cavalleri_nmat14,wangNL_prx20}, whose central frequency is significantly influenced by the compressibility \cite{pimenov_prl01,pimenov_vdm_prl01} due to a capacitive coupling between planes surviving at vanishing momentum \cite{sellati_prb23}. In the basis $\boldsymbol{\psi}_{abc}=\begin{pmatrix}\psi_a & \psi_b &\psi_c\end{pmatrix}^T$, and by taking the limit for infinite compressibility that is appropriate for cuprates,
\begin{align}\lb{effactpsi}
    S[\boldsymbol\psi_{abc}]=\frac{1}{32\pi e^2}\sum_{q}\boldsymbol{\psi}_{abc}^T(-q)\begin{pmatrix}
        \Omega_m^2\varepsilon_{ab}+c^2k_c^2 & 0 & -c^2k_ak_c \\
        0 & \Omega_m^2\varepsilon_{ab}+c^2|\textbf k|^2 & 0 \\
	-c^2k_ak_c & 0& \Omega_m^2\varepsilon_c+c^2k_a^2
    \end{pmatrix}\boldsymbol{\psi}_{abc}(q),
\end{align}
where we have set the in-plane momentum along the $a$-direction ($k_b=0$) without loss of generality, such that $\psi_b$ is decoupled, in full analogy with the case of Eq.\ \pref{jlbt} above.
In the action we have defined, in the Matsubara formalism, the in-plane dielectric function
\begin{align}\lb{inpl}
    \varepsilon_{ab}(i\Omega_m)=\varepsilon_\infty\left(1+\frac{\omega_{ab}^2}{\Omega_m^2}\right),
\end{align}
and the out-of-plane dielectric function
\begin{align}\lb{outpl}
    \varepsilon_{c}(i\Omega_m)=\varepsilon_\infty\left(1+\frac{\omega_{c}^2}{\Omega_m^2}\right),
\end{align}
where the plasma frequencies are linked to the in-plane and out-of-plane superfluid stiffness, $\omega_{ab}^2=4\pi e^2 D_{ab}/\varepsilon_\infty$ and $\omega_c^2=4\pi e^2 D_c/\varepsilon_\infty$ respectively. Indeed, these go back to Eqs.\ \eqref{inpl1} and \eqref{outpl1} once the analytic continuation $i\Omega_m\to \omega+i0^+$ is performed.
Notice that the dielectric tensor is diagonal in the basis $\boldsymbol\psi_{abc}$, as $(a,b,c)$ is the reference frame of the principal axes of the crystal. 
By their definition in Eq.\ \eqref{gauginv}, the gauge invariant fields are formally proportional to currents, and we can then apply within the effective-action framework the same procedure used above for the classical approach, i.e.\ a change of the reference frame to describe a transverse dielectric tensor. We thus perform a rotation around the $b$-axis that combines the $\psi_a$ and $\psi_c$ components into transverse $\psi_t$ and longitudinal $\psi_l$ components with respect to the momentum $\textbf k$. The matrix that performs the change of basis $\boldsymbol{\psi}_{abc}\to\boldsymbol{\psi}_{tbl}=\begin{pmatrix}\psi_t & \psi_b &\psi_l\end{pmatrix}^T$ reads
\begin{align}\lb{rot1}
    U=\begin{pmatrix}
        k_c/|\textbf k| & 0 & k_a/|\textbf k| \\
        0 & 1 & 0 \\
        -k_a/|\textbf k| & 0 & k_c/|\textbf k|
    \end{pmatrix},
\end{align}
and Eq.\ \eqref{effactpsi} transforms in this basis as
\begin{align}\lb{actpsiLT}
    S[\boldsymbol\psi_{tbl}]=\frac{1}{32\pi e^2}\sum_{q}\boldsymbol{\psi}_{tbl}^T(-q)\mathcal D_{tbl}^{-1}\boldsymbol{\psi}_{tbl}(q),
\end{align}
where the matrix of the coefficients reads
\begin{align}\lb{Dlt}
    \mathcal D_{tbl}^{-1}=\begin{pmatrix}
        \Omega_m^2(\varepsilon_{ab}k_c^2+\varepsilon_ck_a^2)/|\textbf k|^2+c^2|\textbf k|^2 & 0 &\Omega_m^2(\varepsilon_c-\varepsilon_{ab})k_ak_c/|\textbf k|^2 \\
        0 & \Omega_m^2 \varepsilon_{ab}+c^2 |\textbf k|^2 & 0 \\
	\Omega_m^2(\varepsilon_c-\varepsilon_{ab})k_ak_c/|\textbf k|^2 & 0 &\Omega_m^2(\varepsilon_{ab}k_a^2+\varepsilon_ck_c^2)/|\textbf k|^2
    \end{pmatrix}.
\end{align}
Before moving forward and studying the linear response, we here provide a brief review of the generalized plasma modes that Eq. \eqref{actpsiLT} describes, useful in the following to provide a physical interpretation of the finite-frequency peak in the real part of the conductivity. The action identifies two longitudinal-transverse mixed modes and one decoupled purely transverse mode along the $b$-direction. The former ones cannot be studied separately, as the anisotropy of layered superconductors is such that the $\psi_t$ and $\psi_l$ components are coupled for generic direction of the momentum, that is, the off-diagonal elements of Eq.\ \eqref{Dlt} are nonvanishing. On physical grounds, this is a manifestation of retardation effects: as already seen in the previous section, at generic wavevector the current induced in the system is not parallel to ${\bf E}$.  This induces a longitudinal electric field in the system in response to a {\em transverse} perturbation, making longitudinal and transverse response unavoidably mixed. Since the displacement current scales as $\pd {\bf E}/\pd (ct)$, the corrections coming from retardation effects are also named relativistic, as they vanish for $c\to \infty$.  The dispersion relations of the two modes obtained from Eq.\ \pref{Dlt} read
\begin{align}\lb{omegapm}
    \omega_\pm^2(\textbf k)=\frac{1}{2}\left(\omega_{ab}^2+\omega_c^2+\frac{c^2}{\varepsilon_\infty}|\textbf k|^2\pm\sqrt{(\omega_{ab}^2-\omega_c^2)^2+\frac{c^4}{\varepsilon_\infty^2}|\textbf k|^4-2\frac{c^2}{\varepsilon_\infty}(k_a^2-k_c^2)(\omega_{ab}^2-\omega_c^2)}\right),
\end{align}
A detailed discussion of the properties of the generalized plasma modes of single-layer anisotropic superconductors can be found in Ref.\ \cite{gabriele_prr22}. Nonetheless, it is important here to stress the main physical outcomes of the present derivation. The generalized dispersions \pref{omegapm} describe two regular functions of the momenta that give $\omega_+(\textbf k\to0)\to\omega_{ab}$ and $\omega_-(\textbf k\to 0)\to\omega_c$. For generic propagation direction $\eta$ and for momenta $|\textbf k| \lesssim \bar k=\sqrt{\varepsilon_\infty(\omega_{ab}^2-\omega_c^2)}/c$ these modes have mixed longitudinal/transverse character, with a degree of mixing that is maximum at $\eta=\pi/4$ and vanishes as one moves along the main crystallographic directions ($k_a=0$ or $k_c=0$), as one immediately realizes by the structure of the off-diagonal matrix elements of Eq.\ \pref{Dlt} scaling as $k_ak_c$. Explicitly neglecting this coupling, i.e.\ setting the off-diagonal elements to zero, would result in having the two modes uncoupled, one of which purely transverse and the other purely longitudinal. In this case the dispersion relation of the latter, that is by definition the plasma mode of the system, can be found by setting to zero the bottom-right element of $\mathcal{D}_{tbl}^{-1}$:
\begin{align}\lb{chareq}
\varepsilon_{ab}(\omega)\frac{k_a^2}{|{\bf k}|^2}+\varepsilon_c(\omega)\frac{k_c^2}{|{\bf k}|^2}=    \varepsilon_{ab}(\omega)\sin^2\eta+\varepsilon_c(\omega)\cos^2\eta=0,
\end{align}
where we have performed the analytic continuation $i\Omega_m\to\omega+i0^+$ and used $k_c=|\textbf k|\cos\eta$ and $k_a=|\textbf k|\sin\eta$.
Using the definitions of the dielectric functions in Eqs.\ \eqref{inpl} and \eqref{outpl}, the solution of Eq.\ \eqref{chareq} is exactly the frequency
\be\lb{omegalk}
\omega_l^2(\bk)=\omega^2_{ab}\frac{k_a^2}{|{\bf k}|^2}+\omega^2_{c}\frac{k_c^2}{|{\bf k}|^2}\equiv
\omega_{ab}^2\sin^2\eta+\omega_c^2\cos^2\eta,
\ee
defined in Eq.\ \eqref{rpadisp}. In addition, one can easily see from Eq.\ \pref{omegapm} that in the limit $c\to \infty$, i.e. in the regime where $\bar k/|\textbf k| \to 0$, retardation (or relativistic) effects can be neglected and one obtains
\be\lb{omegamlim}
\omega_-(\bk)\to  \omega_l(\bk), \quad |\textbf k|\gg \bar k=\sqrt{\varepsilon_\infty(\omega_{ab}^2-\omega_c^2)}/c.
\ee
In other words, the expression $\omega_l(\eta)$ defines the longitudinal-plasmon dispersion in a layered superconductor that one obtains by neglecting retardation effects, as one usually does in the standard RPA approach where only Coulomb interactions are included \cite{randeria_prb00,benfatto_prb01,benfatto_prb04,millis_prr20,dassarma_prl90,dassarma_prb91,dassarma_prb95}. 
We also note in passing that the limit of $\omega_l(\bk)$ for $\textbf k \to 0$ is non-regular as it depends on the direction $\eta$ of the momentum. As shown above, this is not the case for the real electromagnetic mode $\omega_-$, that is regular at $|\textbf k|=0$. In Fig.\ \ref{fig2}(a) we show $\omega_-(\textbf k)$ and $\omega_l(\eta)$ for small values of the propagation angle: as one can see, as $|\textbf k|$ overcomes the $\bar k$ scale $\omega_-$ rapidly approaches the $\omega_l$ limit and the mode becomes longitudinal. By using realistic values of plasma frequencies in cuprates one sees that $\bar k\sim \mu\text{m}^{-1}$. As such, this scale is two orders of magnitude smaller than the momenta usually accessible in RIXS \cite{lee_rixs_nature18,liu_rixs_npjqm20,zhou_prl20,huang_rixs_prb22,hepting_rixs_cm23} or EELS \cite{mitrano_pnas18,mitrano_prx19,abbamonte_natcomm23} experiments, that are not sensitive to the relativistic regime and probe the plasmon dispersion given by Eq.\ \pref{omegalk}. 
\subsection{Interpretation of the conductivity peak of plasmons}
From the action in Eq.\ \eqref{actpsiLT} we can perform the integration of $\psi_l$ and work with an action of the transverse components $\boldsymbol{\psi}_{tb}=\begin{pmatrix}\psi_t & \psi_b\end{pmatrix}^T$ only. This procedure is equivalent to using Ampere's law as a condition to eliminate the longitudinal components, see Eq.\ \eqref{jlbt} and the discussion below. One is left with an action that reads
\begin{align}\lb{actpsiT}
    S[\boldsymbol{\psi}_{tb}]=\frac{1}{32\pi e^2}\sum_q\boldsymbol{\psi}_{tb}^T(-q)\begin{pmatrix}
        \Omega_m^2\varepsilon_t+c^2|\textbf k|^2 & 0 \\ 0 & \Omega_m^2\varepsilon_{ab}+c^2|\textbf k|^2
    \end{pmatrix}\boldsymbol{\psi}_{tb}(q),
\end{align}
%
%
where
\begin{align}\lb{epsitgen}
    \varepsilon_t(i\Omega_m,\eta)=\frac{\varepsilon_{ab}\varepsilon_{c}}{\varepsilon_{ab}\sin^2\eta+\varepsilon_c\cos^2\eta},
\end{align}
is a dielectric function that describes the transverse linear response of the superconductor along the $t$-axis. Indeed, by making use of the relation $\varepsilon_\alpha=\varepsilon_\infty+4\pi i\sigma_\alpha/\omega$ between the optical conductivity and the dielectric function along the direction $\alpha$ \cite{melnyk_prb70} one recovers $\sigma_t$ as in Eq.\ \eqref{sigmat}. 
Remarkably, the denominator of $\varepsilon_t$ can be brought back to the left-hand side of the characteristic equation \eqref{chareq} for the uncoupled longitudinal mode. Indeed, by using the explicit expressions in Eqs.\ \eqref{inpl} and \eqref{outpl} for the in-plane and out-of-plane dielectric functions of plasma modes and performing the analytic continuation $i\Omega_m\to\omega+i0^+$, one can rewrite Eq.\ \eqref{epsitgen} as $\varepsilon_t(\omega,\eta)=\varepsilon_\infty(\omega^2-\omega_{ab}^2)(\omega^2-\omega_c^2)/[\omega^2(\omega^2-\omega_l^2(\eta))]$, exactly as in Eq.\ \pref{epsit1} above.
\begin{figure}[t!]
\centering
\includegraphics[width=14.5cm]{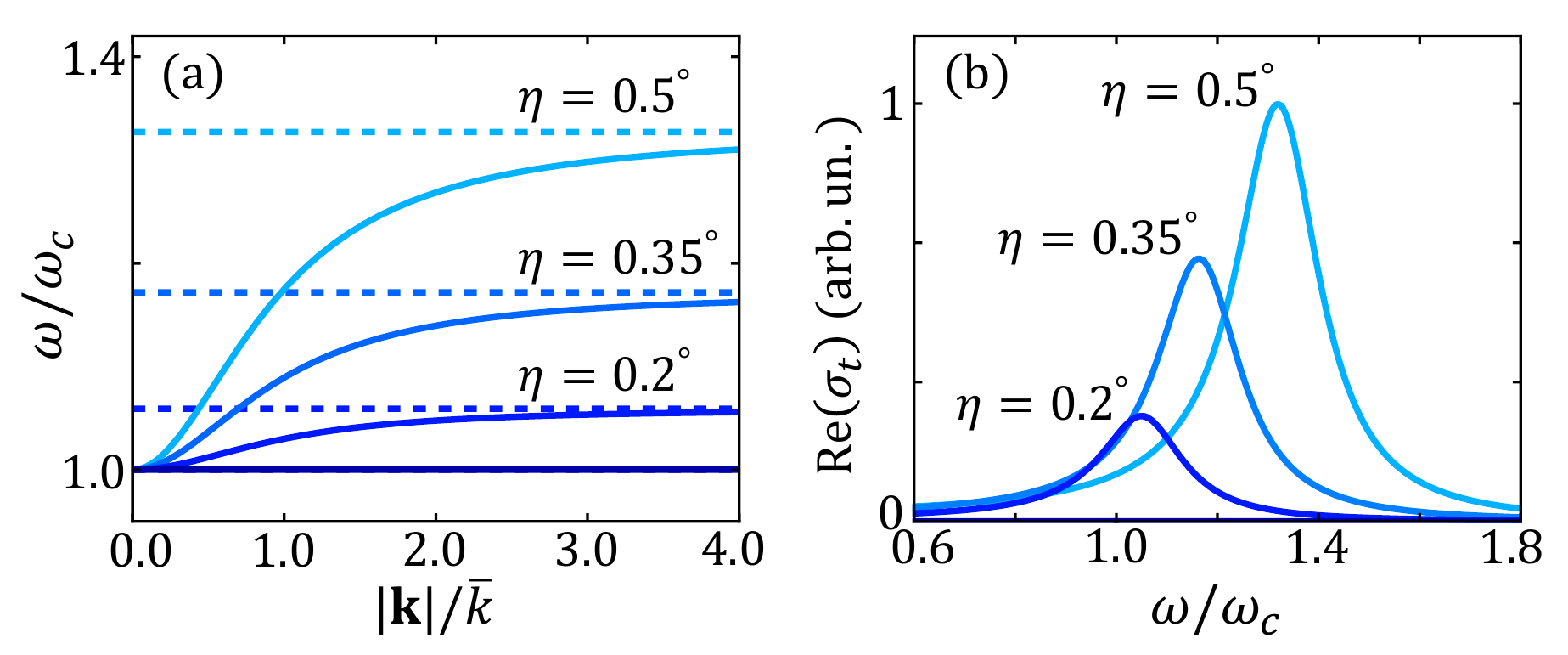}
\caption{\label{fig2} (a) Dispersion of the Josephson plasma mode $\omega_-(\textbf k)$ (solid lines) and $\omega_l(\eta)$ (dashed lines) for different small propagation angles, having chosen $\omega_{ab}/\omega_c=100$. (b) Real part of the conductivity $\sigma_t$ in the case of superconducting plasma modes for corresponding values of $\eta$ of panel (a). The conductivity spectra are normalized to the maximum value of the peak at $\eta=0.5^\circ$. Phenomenological damping parameter is taken as $\gamma=0.1\omega_c$.} 
\end{figure}
While this result has been formally obtained already within the Maxwell's classical formalism of the previous Section, we can now identify the energy of the peak in the {\em transverse} conductivity at vanishing momentum as the value of the {\em longitudinal} plasma mode in the high-momentum regime, i.e.\ the same usually probed by EELS and RIXS, since $\omega_l(\eta)$ is a good approximation of the dispersion of the lower mode $\omega_-(\textbf k)$ for $|\textbf k|\gg \bar k$, see Eq.\ \pref{omegamlim}.\\
The real part of the conductivity $\sigma_t$ is shown in Fig.\ \ref{fig2}(b), where we also introduced a finite damping parameter $\gamma$ when performing the analytic continuation $i\Omega_m\to \omega+i\gamma$. We emphasize once more that such a peak is not a direct manifestation of the Josephson plasmon of the superconductor \cite{pimenov_prb00,pimenov_apl00,pimenov_prb02,armitage_prb21} which, as discussed above, for vanishing momentum is at frequency $\omega_c$ for every direction $\eta$. Indeed, plasma modes appear as zeroes of the dielectric function and do not lead to finite-frequency peaks in the conductivity. Instead, the absorptive peak at $\omega_l(\eta)$ is a manifestation of the mixing mechanism between in-plane and out-of-plane plasma modes described in the previous section, as the dielectric function in Eq.\ \eqref{epsit1} comes directly from the action for the coupled modes Eq.\ \eqref{actpsiLT}. On a more general ground, our derivation clarifies that a signature of longitudinal nature appears in the transverse response whenever the longitudinal mode is coupled to the transverse one without directly participating in the detection, that is, the degree of freedom is integrated out.\\
It is worth mentioning that our derivation is not restricted to electron-doped cuprates, in which the peak has already been experimentally reported \cite{pimenov_prb00,pimenov_apl00,pimenov_prb02,armitage_prb21}, but it is in principle valid for any single-layer superconductor, like the hole-doped LSCO. We also point out that the results could be extended to bilayer superconductors like YBCO, that display two Josephson plasmons at frequencies $\omega_{c1}$ and $\omega_{c2}$. Indeed, by using the out-of-plane bilayer dielectric function \cite{vandermarel96} $\varepsilon_c=\varepsilon_\infty(\omega^2-\omega_{c1}^2)(\omega^2-\omega_{c2}^2)/[\omega^2(\omega^2-\omega_T^2)]$, with $\omega_T^2=\omega_{c1}^2d_2+\omega_{c2}^2d_1$ and $d_{1,2}$ the intra- and inter-bilayer spacings, one predicts two absorptive peaks in the conductivity centered at the high-momenta values of the dispersions of the Josephson modes \cite{sellati_prb23}. The high-energy one follows the same trend of the peak in single-layer superconductors, moving with $\eta$ from $\omega_{c1}$ to $\omega_{ab}$. The low-energy one quickly moves from $\omega_{c2}$ to $\omega_T$ even for small values of $\eta$, and does not disappear for $\eta=\pi/2$ \cite{homes_prl93,vandermarel96,yamada_prl98,erb_prl00,vandermarel_prb01,bernhard_prl11,cavalleri_nmat14,wangNL_prx20}. As mentioned above, finite compressibility corrections are crucial for optical measurements in bilayer superconductors \cite{pimenov_prl01,pimenov_vdm_prl01}, and they must be taken into account to correctly fit the experimental data \cite{sellati_prb23}.

\section{Fresnel equations at normal incidence on a tilted-grown sample}
To link the results obtained in the previous sections to experiments we must consider the measured quantity, that is the electric field transmitted or reflected through the sample with respect to the incident wave, and link it to the conductivity $\sigma_t$. Moreover, one might argue that due to the fact that the system is anisotropic, both angles $\eta$ and $\phi_\text J$ that define the current propagation within the material differ respectively from $\eta_{in}$, the angle between the incident momentum of the external wave and the normal to the planes, and $\phi$, angle between the $b$-axis and the electric field that describes its polarization. To this aim, we must write the Fresnel conditions at the boundaries of the sample. 
In this section we analyze the configuration in which a THz pulse is at normal incidence on a thin-film layered superconductor, grown with tilted planes at a small angle $\eta$ \cite{pimenov_prb00,pimenov_apl00,pimenov_prb02,armitage_prb21}, and we show that in this case the theoretical results can be easily related to the experiments, see Appendix A.
\begin{figure}[t!]
\centering
\includegraphics[width=18.5cm]{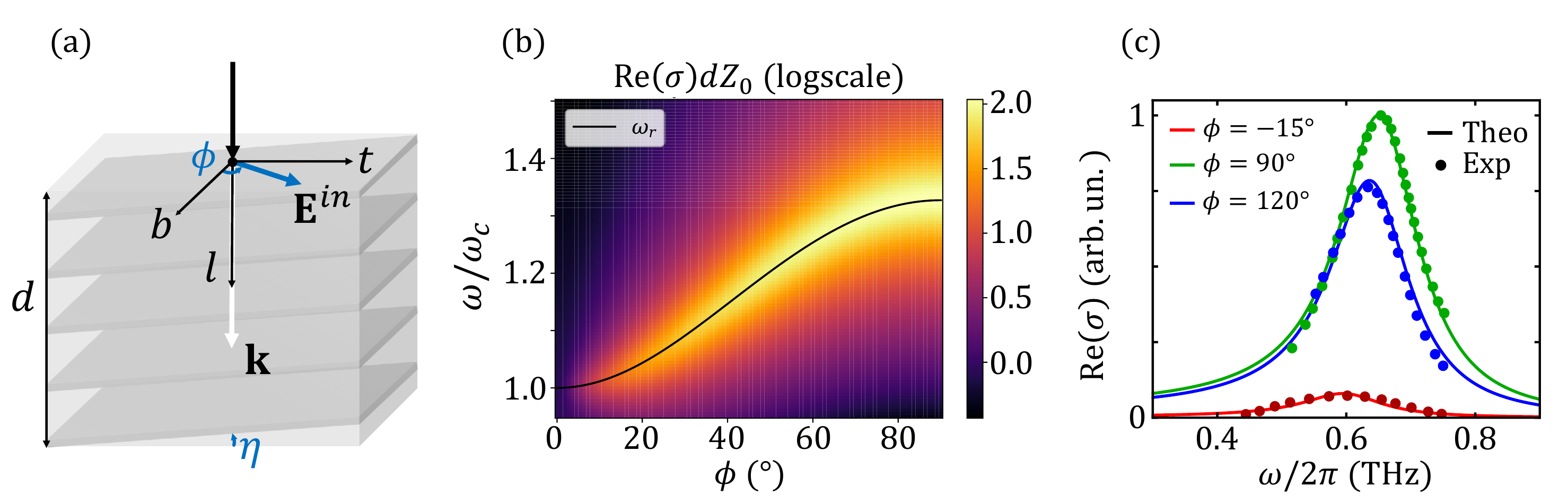}
\caption{\label{fig3} (a) Sketch of the experimental configuration with a THz wave at normal incidence on a tilted-grown sample of thickness $d$. The reference frame $(t,b,l)$ for this configuration is also shown, to highlight the direction of the external electric field $\textbf E^{in}$ that defines the polarization angle $\phi$. (b) Real part of the measured conductivity as a function of frequency and polarization angle as in Eq. \eqref{sigfromt}. Solid black line corresponds to $\omega_r(\eta,\phi)$ as in Eq. \eqref{sigplas}. In this plot $\eta=0.25^\circ$, $d=0.150\text{ }\mu\text m$, $\omega_{ab}/\omega_c=100$ and $\gamma=0.1 \omega_c$. (c) Fit of experimental data from Ref.\ \cite{armitage_prb21} with $\sigma(\omega,\eta,\phi)$ for different polarization angles. Fitting parameters are extracted at once from the three measurements: $\omega_{ab}/2\pi=60\text{ THz}$, $\omega_c/2\pi=0.6\text{ THz}$, $\eta=0.26^\circ$, $\gamma=0.075\text{ THz}$.}
\end{figure}\\
Following the notation set above, we define the reference frame $(t,b,l)$ in such a way that the $tb$-plane corresponds to the interface and the $l$-axis is perpendicular to it, see Fig.\ \ref{fig3}(a). At normal incidence $\eta_{in}=\eta$ immediately, as the momentum of the wave does not change its direction while crossing the interface. Within the material, the $b$-polarized and $t$-polarized electric fields are decoupled and travel with different values of the wave-vector, see Eq.\ \eqref{actpsiLT} and the discussion below. In particular, from Eq.\ \eqref{Dlt} the equations of motion read $|\textbf k|^2=\omega^2\varepsilon_{ab}/c^2$ for the former and $|\textbf k|^2=\omega^2\varepsilon_t/c^2$ for the latter \cite{bulaevskii_prb02}, with $\varepsilon_t$ defined in Eq.\ \eqref{epsitgen}. We then impose the continuity of the tangential components of the electric field, $\text E_t$ and $\text E_b$, and of the magnetic field, $\text B_t$ and $\text B_b$, at the interface $l=0$ and at $l=d$, with $d$ the sample thickness. By solving the system set by these conditions one finds the transmission and reflection coefficients for the $t$ and the $b$ components of the field in the thin-film configuration, that read
\begin{align}
    \lb{Tt}& \text T_t=\frac{\mathcal T_t\mathcal T_t^{\prime}e^{in_t\omega d/c}}{1-\mathcal R_t^2e^{2in_t\omega d/c}},\\
    \lb{Rt}& \text R_t= \frac{\mathcal R_t(1-e^{2in_t\omega d/c})}{1-\mathcal R_t^2e^{2in_t\omega d/c}},\\
     \lb{Tb}& \text T_b=\frac{\mathcal T_b\mathcal T_b^{\prime}e^{in_b\omega d/c}}{1-\mathcal R_b^2e^{2in_b\omega d/c}},\\
    \lb{Rb}& \text R_b= \frac{\mathcal R_b(1-e^{2in_b\omega d/c})}{1-\mathcal R_b^2e^{2in_b\omega d/c}},
\end{align}
where $n_\alpha=\sqrt{\varepsilon_\alpha}$ is the refractive index along the direction $\alpha$, $\mathcal{T}_\alpha=2/(1+n_\alpha)$ is the transmission coefficient going from vacuum to the material, $\mathcal{T}^\prime_\alpha=2n_\alpha/(1+n_\alpha)$ is analogously the transmission coefficient from the sample to the vacuum, and $\mathcal{R}_\alpha^2=1-\mathcal{T}_\alpha\mathcal{T}_\alpha^\prime$ accounts for the Fabry-Perot interference within the thin film. The ratios $\text T_t/\text T_b$ and $\text R_t/\text R_b$ carry the information on the rotation of the polarization of the transmitted or reflected wave. By the definition of the dielectric function $\varepsilon_t$ in Eq. \eqref{epsitgen} one has that $\varepsilon_t\simeq\varepsilon_{ab}$ under the assumption of small tilt angle of the planes. Then $n_t\simeq n_{b}$ and the ratios are approximately 1: one can thus conclude that the polarization of the transmitted or reflected wave does not differ significantly from the one of the incident wave in the experiment. With the same reasoning $\sigma_t\simeq \sigma_{ab}$, so that the transverse current is approximately parallel to the field, see Eq.\ \eqref{jbt}, and we can thus conclude that $\phi \simeq \phi_\text J$. 
In an experiment the measured quantity (see Appendix A) is either the transmissivity
\begin{align}\lb{ttot}
\text T=\text T_b \cos^2 \phi+\text T_t \sin^2 \phi,
\end{align}
or, analogously, the reflectivity
\begin{align}\lb{rtot}
\text R=\text R_b \cos^2 \phi+\text R_t \sin^2 \phi.
\end{align}
From these quantities one can define the measured transverse conductivity. Indeed, under the assumption of film-thickness $d$ much smaller than the wavelength of the radiation inside the material and its penetration depth, one finds \cite{armitage_prb21}
%
\begin{align}\lb{sigfromt}
    \sigma(\omega,\eta,\phi)=\frac{2}{Z_0d}\left(\frac{1}{\text T}-1\right),
\end{align}
where $Z_0=4\pi/c$ is the impedance of free space. This proportionality establishes the link between the measured quantity and the theoretical conductivity we were looking for. Moreover, in the case of cuprates one can numerically estimate $\text T\ll 1$ in Eq.\ \eqref{ttot}, and then approximate $\sigma\propto1/\text T$. Since also $\sigma_{ab}\propto1/\text T_b$ and $\sigma_t\propto1/\text T_t$, from \eqref{ttot} one can express the measured transverse conductivity as
\begin{align}\lb{sigmatau}
    \sigma(\omega,\eta,\phi)\simeq\frac{\sigma_{ab}\sigma_t}{\sigma_{ab}\sin^2\phi+\sigma_t\cos^2\phi}.
\end{align}
With $\sigma_t$ from Eq.\ \eqref{sigmat} and using the expressions of $\sigma_{ab}$ and $\sigma_c$ for the superconductor, one finds that the real part of the conductivity has a peak at a resonance frequency $\omega_r(\eta,\phi)$ that depends on both $\eta$ and $\phi$:
\begin{align}\lb{sigplas}   
    \omega_r^2(\eta,\phi)=\frac{\omega_{ab}^2\sin^2\eta\sin^2\phi+\omega_c^2(1-\sin^2\eta\sin^2\phi)}{1-\sin^2\eta\cos^2\phi+\left(\frac{\omega_c}{\omega_{ab}}\right)^2\sin^2\eta\cos^2\phi}.
\end{align}
In Fig.\ \ref{fig3}(b) we show the real part of Eq.\ \eqref{sigfromt} as a function of the external polarization angle and we compare the peak emerging in the measured conductivity with Eq.\ \eqref{sigplas}.  Indeed, the approximated expression \eqref{sigmatau} provides an excellent description of the experimental data in Refs.\ \cite{pimenov_prb00,pimenov_apl00,pimenov_prb02,armitage_prb21}, and the frequency Eq.\ \eqref{sigplas} establishes a link between the peak of the experimental conductivity and the plasma frequencies $\omega_{ab}$ and $\omega_{c}$, which can then be extracted as fitting parameters given the angles $\eta$ and $\phi$. In Fig.\ \ref{fig3}(c) we fit experimental data from Ref.\ \cite{armitage_prb21} to provide an estimate of the in-plane and out-of-plane plasma frequencies of the overdoped $\text{La}_{1.87}\text{Ce}_{0.13}\text{CuO}_4$ ($T_c=21$ K) at $5$ K. We here clarify that the experimental fit is not meant to draw any conclusion on the symmetry of the superconducting order parameter, that is still debated in the context of electron-doped cuprates \cite{armitage_prb21,song_nsr21}. Our starting point Eq.\ \eqref{supfluid} can be derived from a microscopic model which admits a modulation of the superconducting gap \cite{sellati_prb23}. At the level of Eq.\ \eqref{supfluid}, the gap symmetry enters in the temperature dependence of the superfluid stiffness $D_{ab}$ and $D_c$, that is controlled by quasiparticle excitations, while it barely affects the charge compressibility $\kappa_0$. In addition, the gap symmetry can affect the quasiparticle damping of the plasmon, controlling the phenomenological broadening $\gamma$, even though other mechanism can determine its value independently of the gap symmetry. However, once the proper gap symmetry has been embedded in the plasma frequencies $\omega_{ab}$ and $\omega_c$, the structure of the modes Eq.\ \eqref{omegapm} is general.\\
So far, it was only empirically observed in Ref.\ \cite{pimenov_apl00} that the data could be well fitted by using an effective conductivity $\sigma_t(\omega,\eta_\text{eff})$ having the same functional form of Eq.\ \eqref{sigmat}, but with an effective tilt angle $\eta_{\text{eff}}=\eta\sin\phi$. This result actually follows from Eq.\ \pref{sigmatau} in the case of small angle $\eta$ between the momentum and the $c$-axis of the crystal, which is indeed the configuration of Ref.\ \cite{pimenov_apl00}. In this case, the frequency of the peak in Eq.\ \eqref{sigplas} can be approximated as
\begin{align}\lb{sigplaseff}
    \omega_r^2(\eta,\phi)\simeq\omega_{ab}^2\sin^2\eta_\text{eff}+\omega_c^2\cos^2\eta_\text{eff}=\omega_l^2(\eta_\text{eff}),
\end{align}
where again $\eta_\text{eff}=\eta\sin\phi$.\\
At first, one might as well consider the configuration in which the THz pulse is incident with a small angle on a $c$-axis grown sample. However, computing the Fresnel conditions in this case results in featureless transmissivity and reflectivity, and no peak appears in the real part of the conductivity (see Appendix A for details)


\section{Conclusions}
In this manuscript we studied the optical absorption in layered superconductors in a tilted geometry, where the light propagates inside the sample by forming a small angle with the stacking direction. We showed that such a geometry makes it possible to observe with optics, that is essentially a zero-momentum probe, a direct signature of the plasmon dispersion at momenta of the order of a fraction of the Brillouin zone, that is usually probed by RIXS or EELS. The basic physical mechanism behind this observation is the intrinsic mixing between transverse and longitudinal electromagnetic modes in a layered material, due to the anisotropy between the in-plane and out-of-plane response. Such mixing, that is absent when light propagates along the main crystallographic axes, leads to the emergence of an absorption peak in the transverse optical conductivity in tilted geometry. Interestingly, we can show analytically that the peak frequency moves as a function of the tilting angle according to the functional law that the physical longitudinal plasmon displays at momenta larger than the scale where transverse/longitudinal mixing is relevant. In cuprates, where the SC c-axis plasmon is weakly affected by Landau damping due to the opening of a large spectral gap below $T_c$, the peak is well defined at small tilting angle, and it has been indeed observed in several electron-doped cuprates \cite{pimenov_prb00,pimenov_apl00,pimenov_prb02,armitage_prb21}. Here we argue that the same effect can be seen in any layered sample, provided that the appropriate Fresnel geometry is implemented. In addition, we provide an analytical expression for the peak frequency as a function of both tilting angle and light polarization, that can be used to obtain from a single-set of measurements the relevant scales for plasma excitations in these systems. It is worth stressing that in the last few years, after charged plasmons have been detected for the first time with high-resolution RIXS \cite{lee_rixs_nature18,liu_rixs_npjqm20,zhou_prl20,huang_rixs_prb22,hepting_rixs_cm23} and EELS \cite{mitrano_pnas18,mitrano_prx19,abbamonte_natcomm23} experiments, an intense discussion emerged on the nature of charge fluctuations in these correlated materials \cite{mitrano_prx19,abbamonte_natcomm23}. The all-optical measurement proposed here is in principle a bulk probe, it is not affected by the lack of sensitivity at small momenta connected to plasmon measurements via charge-detecting probes, and it allows for a precise control on the momentum value, that can be problematic e.g.\ to EELS \cite{abbamonte_natcomm23}. As a consequence, the experimental verification of this idea could provide an additional knob to explore charge fluctuations in cuprates, and their possible interplay with other collective modes of the systems.

\section*{Acknowledgements}
We acknowledge useful discussion with N.P. Armitage. We acknowledge financial support by EU under project MORE-TEM ERC-Syn grant agreement No. 951215 and by Sapienza University of Rome under Project Ateneo 2022 RP1221816662A977 and Project Ateneo 2023 RM123188E357C540.

\section*{Appendix A}
\begin{figure}[h!]
\centering
\includegraphics[width=11.cm]{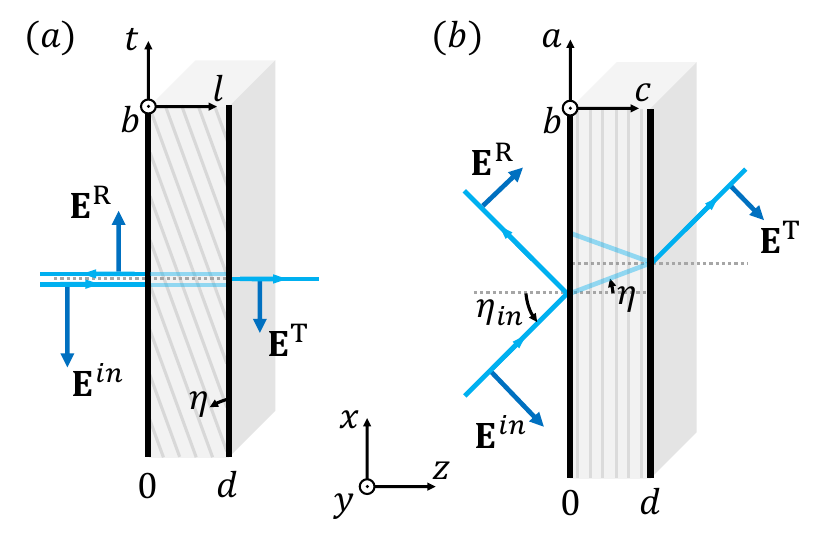}
\caption{\label{fig_a1} Sketch of the two possible experimental configurations discussed in this Appendix. In both cases the reference frame $(x,y,z)$ is defined such that $xy$ is the vacuum-sample interface and $xz$ is the plane of incidence. For graphical purposes, only TM polarized waves are depicted. (a) Geometry of the configuration in which the incident THz pulse is at normal incidence on a thin-film crystal grown with layers tilted at angle $\eta$ with respect to the vacuum-sample interface, as in Section 3.3. In this case the $(x,y,z)$ reference frame corresponds to the $(t,b,l)$ frame introduced in the main text. (b) Geometry of the configuration in which the incident field is at oblique incidence with angle $\eta_{in}$ on a thin-film sample with planes parallel to the interface. In this case the $(x,y,z)$ reference frame corresponds to the crystallographic $(a,b,c)$ frame.}
\end{figure}

\begin{figure}[t!]
\centering
\includegraphics[width=14.cm]{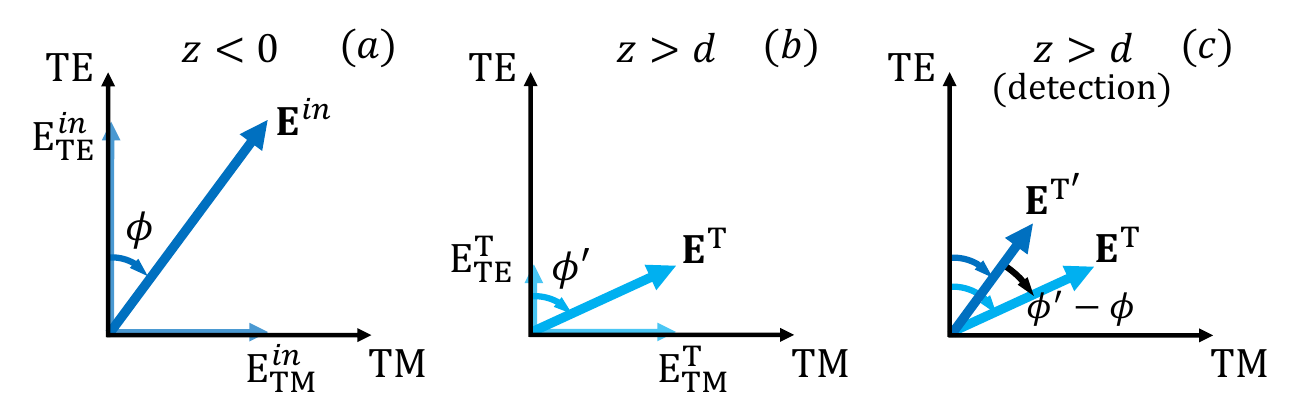}
\caption{\label{fig_a2} (a) Incoming electric field $\mathbf{E}^{in}$ at a generic polarization angle $\phi$ with respect to the TE axis (corresponding to the $y$ axis and to the crystallographic $b$ axis in both configurations of Fig.\ \ref{fig_a1}). (b) Transmitted electric field $\mathbf{E}^\text T$ found by means of the appropriate set of boundary conditions. The transmitted field still lies in the TM-TE plane due to conservation of the parallel component of the incident momentum, but $\mathbf{E}^\text{T}$ is generically polarized with angle $\phi^{\prime}\neq\phi$. (c) Comparison between the directions of the transmitted field $\textbf E^\text T$ and of the projected field $\textbf E^{\text{T}^\prime}$ along the incoming polarization, that is the measured quantity in a transmission experiment.}
\end{figure}

In this Appendix we derive the transmitted electric field in two experimental configurations in which the electromagnetic wave travels with a finite angle with respect to the stacking direction of the planes,
by means of standard Fresnel-like boundary conditions applied on a uniaxial film. 
Let us consider a transmission experiment on a superconductor placed in the region $0<z<d$, as in Fig.\ \ref{fig_a1}. The electric field satisfies Maxwell's equations
\begin{align}\lb{max}
\begin{cases}
\nabla^2\mathbf{E}-\frac{1}{c^2}\frac{\partial^2\mathbf{E}}{\partial t^2}=0 &z<0,\text{ }z>d\\
\nabla^2\mathbf{E}-\nabla(\nabla\cdot\mathbf{E})-\frac{1}{c^2}\frac{\partial^2(\hat{\boldsymbol\varepsilon}\mathbf{E})}{\partial t^2}=0&0<z<d\\
\end{cases},
\end{align}
where $\hat{\boldsymbol\varepsilon}$ is the dielectric tensor of the uniaxial material. By expanding the electric field on a basis of plane waves, Eq.\ \eqref{max} becomes a linear system for the Cartesian Fourier components of the electric field
\begin{align}
\lb{ap:max}
\begin{cases}
(\lvert\mathbf{k}\rvert^2-\omega^2/c^2)\delta_{\alpha\beta}\text E_{\beta}=0 &z<0,\text{ }z>d\\
(\lvert\mathbf{k}\rvert^2\delta_{\alpha\beta}-k_{\alpha}k_{\beta}-\omega^2\varepsilon_{\alpha\beta}/c^2)\text E_{\beta}=0&0<z<d\\
\end{cases}.
\end{align}
A propagating solution is allowed whenever the determinant of this system
is zero. 
For the experimental configuration of Fig.\ \ref{fig_a1}(a) in which the frame $(x,y,z)$ corresponds to $(t,b,l)$, as in Section 3.3,
the incoming momentum is along the $z$ (or $l$) direction.
In this geometry the dielectric tensor is defined as (see Eqs.\ \eqref{jlbt} and \eqref{Dlt})
\begin{align}
\hat{\boldsymbol\varepsilon}=
\begin{pmatrix}
\varepsilon_{ab}\cos^2{\eta}+\varepsilon_{c}\sin^2{\eta} & 0 & (\varepsilon_{ab}-\varepsilon_{c})\cos{\eta}\sin{\eta} \\
0 & \varepsilon_{ab} & 0 \\
(\varepsilon_{ab}-\varepsilon_{c})\cos{\eta}\sin{\eta} & 0 & \varepsilon_{ab}\sin^2{\eta}+\varepsilon_{c}\cos^2{\eta} \\
\end{pmatrix}.
\end{align}
From Eq.\ \eqref{ap:max} one can notice that the subspace associated with $\text E_y$ is decoupled from the one associated with $\text E_x$ and $\text E_z$. As a consequence, one can show immediately that $y$-polarized electric fields in the material propagate with wave-vector $\lvert\mathbf{k}\rvert^2=\omega^2\varepsilon_{ab}/c^2$, while $x$- and $z$-polarized electric fields propagate with wave-vector $\lvert\mathbf{k}\rvert^2=\omega^2\varepsilon_{t}/c^2$, where
\begin{align}
\varepsilon_t=\frac{\varepsilon_{ab}\varepsilon_{c}}{\varepsilon_{ab}\sin^2{\eta}+\varepsilon_{c}\cos^2{\eta}},
\end{align}
as in Eq.\ \eqref{epsitgen} in the main text. Conversely, for the experimental configuration of Fig.\ \ref{fig_a1}(b) in which the frame $(x,y,z)$ corresponds to the crystallographic frame $(a,b,c)$, the dielectric tensor is diagonal $\hat{\varepsilon}_{\alpha\beta}=\varepsilon_{\alpha}\delta_{\alpha\beta}$ (see Eqs.\ \eqref{jabc} and \eqref{effactpsi}). As the wave-vector belongs to the $xz$-plane, again in this case the equation for $\text E_y$ is decoupled from the other two components.
One can then in both cases solve the system separately for the transmission of the $y-$ and for the mixed $xz-$ polarized components of $\mathbf{E}$. In the following we will refer to the former component as the Transverse Electric (TE) field and to the latter as the Transverse Magnetic (TM) field, as one would commonly do in the oblique incidence configuration in which the plane of incidence is $xz$. Notice that the transmitted wave $\textbf E^\text T$ is generically polarized along a direction $\phi^\prime$ that differ from the polarization $\phi$ of the incident wave $\textbf E^{in}$, although the experiment is still set so to measure the outgoing $\phi$-polarized wave $\textbf E^{\text T^\prime}$. Assuming that the TE and TM transmission coefficients are known, such that $\text E^\text{T}_\text{TE,TM}=\text T_\text{TE,TM}\text E^{in}_\text{TE,TM}$ (see Fig. \ref{fig_a2} for the notation), one finds the measured transmitted field polarized along $\phi$ as
%
\begin{align}
	\label{ap:ttot}
\text E^{\text{T}^\prime}&=\text E^\text{T}\cos{(\phi^{\prime}-\phi)}\nn\\
&=\text E^\text{T}\cos{\phi^{\prime}}\cos{\phi}+\text E^\text{T}\sin{\phi^{\prime}}\sin{\phi}\nn\\
&=\text T_\text{TE}\text E^{in}_\text{TE}\cos{\phi}+\text T_\text{TM}\text E^{in}_\text{TM}\sin{\phi}\nn\\
&=(\text T_\text{TE}\cos^2{\phi}+\text T_\text{TM}\sin^2{\phi})\text E^{in}\equiv \text T \text E^{in}.
\end{align}
Analogously, one can express the reflected field in a similar way. For the configuration discussed in the Section 3.3, see Fig.\ \ref{fig_a1}(a), the TE and TM components stand for the $b$ and $t$ components respectively, and one recovers Eq.\ \eqref{ttot} of the main text. We now compute explicitly the transmission coefficients $\text T_\text{TE}$ and $\text T_\text{TM}$ for the two configurations separately.\\
For the first configuration (Fig. \ref{fig_a1}(a)) $\text T_\text{TE}\equiv \text T_b$ and $\text T_\text{TM}\equiv \text T_t$, thus the TE and TM components propagate with different refractive indices, $n_{b}=\sqrt{\varepsilon_{ab}}$ and $n_t=\sqrt{\varepsilon_t}$ respectively. Imposing the continuity of the tangential components of the electric \textbf E and magnetic \textbf B fields, one recovers the usual expression for transmission at normal incidence on a slab, namely
\begin{equation}\lb{talpha}
	\text T_\alpha=\frac{\mathcal T_\alpha \mathcal T^{\prime}_\alpha e^{in_\alpha\omega d/c}}{1-\mathcal R_\alpha^2e^{2in_\alpha\omega d/c}},
\end{equation}
where $\mathcal T_\alpha=2/(n_{\alpha}+1)$ is the transmission coefficient from vacuum $n=1$ to a medium with refractive index $n_\alpha$, $\mathcal T^{\prime}_\alpha=2n_\alpha/(n_\alpha+1)$ is analogously the transmission coefficient from medium to vacuum and the denominator accounts for the Fabry-Perot interference inside the slab of thickness $d$, with $\mathcal R^2_\alpha=1-\mathcal T_\alpha \mathcal T_\alpha^{\prime}$. Eq.\ \eqref{talpha} can also be expressed as
\begin{equation}
	\text T_\alpha=\frac{2n_\alpha}{2n_\alpha\cos{\zeta_\alpha}-i(n_\alpha^2+1)\sin{\zeta_\alpha}},
\end{equation}
where $\zeta_\alpha=n_\alpha\omega d/c$. In the approximation $d\ll\lvert n_\alpha\rvert\omega/c$, i.e.\ the thickness of the film is much smaller than the wavelength of the radiation inside the material $\lambda=\text{Re}(n_\alpha)\omega/c$ and the skin depth $\delta=\text{Im}(n_\alpha)\omega/c$, one can take at first order in $\zeta_\alpha$ 
\begin{equation}
	\label{ap:approx}
	\frac{1}{\text T_\alpha}\approx1-i\frac{n_\alpha^2+1}{2n_\alpha}\zeta_\alpha=1-i\frac{(n_\alpha^2+1)\omega d}{2c}.
\end{equation}
Using the relation between refractive index and conductivity $\sigma_\alpha=\frac{\omega}{4\pi i}(n_\alpha^2-1)$ one can rewrite Eq.\ \eqref{ap:approx} as
\begin{equation}
	\frac{1}{\text T_\alpha}=1+\frac{4\pi d\sigma_\alpha}{2c}-i\frac{2\omega d}{c}\approx1+\frac{4\pi d\sigma_\alpha}{2c},
\end{equation}
where again we considered $\omega d/c\ll1$. Consequently, one finds
\begin{equation}\lb{sigmavst}
	\sigma_\alpha=\frac{2}{Z_{0}d}\left(\frac{1}{\text T_\alpha}-1\right),
\end{equation}
where $Z_{0}=4\pi/c$. This relation between the conductivity and the transmissivity is valid along both the $b$ and $t$ directions. On the other hand, one can imagine to extract an experimental conductivity from the experimental transmissivity $\text T$ as in Eq.\ \eqref{ap:ttot} by applying the same relation, see e.g.\ Ref.\ \cite{armitage_prb21} where Eq.\ \eqref{sigfromt} is used. \\ 
In the second configuration (Fig.\ \ref{fig_a1}(b)) the interface is parallel to the $ab$-plane of the crystal and one needs to solve a wider set of continuity conditions. Indeed, one must impose the continuity of the tangential components of the electric $\textbf E$ and magnetic $\textbf B$ fields as for the previous configuration, but also the continuity of the normal component of the displacement field $\textbf D$ \cite{wooten_fresnel68,menovsky_prb97}.  
To understand how the transmission occurs in this case, let us first recall the results expected for an isotropic film, where the propagation of the electromagnetic wave inside the sample is defined by a unique refractive index $n$. In this case one easily finds that $\text T^\text{(iso)}\equiv \text{E}^{\text T^\prime} / \text{E}^{in}$ reads:
%
\begin{equation}\lb{isot}
	\text T^\text{(iso)}(\eta_{in},n)=\frac{\mathcal T \mathcal T^{\prime}e^{in\omega d\cos{\eta}/c}}{1-\mathcal R^2e^{2in\omega d\cos{\eta}/c}},
\end{equation}
where $\eta$ is the propagation angle inside the material, see Fig.\ \ref{fig_a1}(b), while $\eta_{in}$ is the external angle of incidence. Here we defined as before $\mathcal T=2\cos{\eta_{in}}/(n\cos{\eta_{in}}+\cos\eta)$ as the transmission coefficient from vacuum to the medium and analogously $\mathcal T^{\prime}=2n\cos{\eta}/(n\cos{\eta_{in}}+\cos{\eta})$ as the transmission coefficient from medium to vacuum, while again the denominator of Eq.\ \eqref{isot} accounts for the Fabry-Perot interference inside the slab of thickness $d$, with $\mathcal R^2=1-\mathcal T\mathcal T^{\prime}$.
In Eq.\ \eqref{isot} we made explicit the dependence of the transmissivity $\text T$ on the incident angle and on the refractive index $n$ only. Indeed, the propagation angle inside the sample is automatically defined by these two quantities thanks to the Snell's law, which states that  $\sin\eta=\sin{\eta_{in}}/n$ (notice that in the isotropic case one has to define $\eta$ with respect to the normal to the interface as there are no planes, but we maintain the notation to press the analogy between the two cases). However, in the uniaxial case of Fig.\ \ref{fig_a1}(b) Snell's relation is not valid, since due to anisotropy of the refractive indices the components $k_{ab}$ and $k_c$ of the momentum are rescaled differently. This makes the transmission coefficient a function of $\eta_{in}, n_b$ and $n_c$. More specifically, for the TM wave one finds that $\text T_\text {TM}$ has an expression analogous to Eq.\ \eqref{isot}, provided that one replaces $n\to n_b=\sqrt{\varepsilon_{ab}}$ and $\sin\eta\to \sin\eta_{in}/n_c=\sin\eta_{in}/\sqrt{\varepsilon_c}$:
\begin{equation}\lb{anisot}
	\text T_\text {TM}(\eta_{in},n_b, n_c)=\frac{\mathcal T \mathcal T^{\prime}e^{in_b\omega d\cos{\eta}/c}}{1-\mathcal R^2e^{2in_b\omega d\cos{\eta}/c}},
\end{equation}
with $\mathcal T$ and $\mathcal T^{\prime}$ retaining the same functional dependence on $\eta, \eta_{in}$ as before.
In this situation, the argument of the complex exponential in Eq.\ \eqref{anisot} reads:
\begin{equation}
	\zeta=n_b\frac{\omega d}{c}\cos{\eta}=n_{b}\frac{\omega d}{c} \sqrt{1-\frac{\sin^2\eta_{in}}{n_c^2}}.
\end{equation}
One can check that for THz frequencies around the Josephson plasma frequency $\omega\sim\omega_c$ the divergence in the square root possible for $\eta_{in}\neq0$ is weakened by the residual quasiparticle damping $\gamma$, so that we obtain $\zeta\ll1$. Then, evaluating $1/\text T_{\text{TM}}$ from \eqref{anisot} at small $\zeta$ one gets

%
\begin{equation}\lb{TM2}
	\frac{1}{\text T_\text{TM}}\approx 1-i\frac{\omega d}{2c}\frac{\varepsilon_{ab}\cos^2{\eta_{in}}+1-\sin^2{\eta_{in}}/\varepsilon_c}{\cos{\eta_{in}}}.
\end{equation}
Similar reasonings can be made for the TE component, that is expressed as in Eq.\ \pref{anisot}, provided that $\sin\eta\to \sin\eta_{in}/n_b=\sin\eta_{in}/\sqrt{\varepsilon_{ab}}$. Also in this case one can approximate the transmission coefficient along this direction as:
\begin{equation}\lb{TE2}
	\frac{1}{\text T_\text {TE}}\approx1-i\frac{\omega d}{2c}\frac{\varepsilon_{ab}\cos^2{\eta_{in}}+1-\sin^2{\eta_{in}}/\varepsilon_{ab}}{\cos{\eta_{in}}}.
\end{equation}
Even though Eqs.\ \eqref{TM2} and \eqref{TE2} still depend on a combination of $\varepsilon_{ab}$ and $\varepsilon_c$, these structures do not lead to the pole observed in the transverse dielectric function $\varepsilon_t$, as opposed to Eq.\ \eqref{ap:approx} obtained in the first configuration. In the end, by explicit numerical computation with realistic parameter values for cuprates of the transmissivity $\text T_\text {TM}$ and $\text T_\text {TE}$ in Eq.\ \pref{anisot}, with the corresponding definitions of $\eta$, we verified that the corresponding conductivities, expressed as in Eq.\  \pref{sigmavst},   are featureless, and no finite-frequency peaks are observed.

\bibliography{bibl.bib}

\begin{thebibliography}{54}%
\makeatletter
\providecommand \@ifxundefined [1]{%
 \@ifx{#1\undefined}
}%
\providecommand \@ifnum [1]{%
 \ifnum #1\expandafter \@firstoftwo
 \else \expandafter \@secondoftwo
 \fi
}%
\providecommand \@ifx [1]{%
 \ifx #1\expandafter \@firstoftwo
 \else \expandafter \@secondoftwo
 \fi
}%
\providecommand \natexlab [1]{#1}%
\providecommand \enquote  [1]{``#1''}%
\providecommand \bibnamefont  [1]{#1}%
\providecommand \bibfnamefont [1]{#1}%
\providecommand \citenamefont [1]{#1}%
\providecommand \href@noop [0]{\@secondoftwo}%
\providecommand \href [0]{\begingroup \@sanitize@url \@href}%
\providecommand \@href[1]{\@@startlink{#1}\@@href}%
\providecommand \@@href[1]{\endgroup#1\@@endlink}%
\providecommand \@sanitize@url [0]{\catcode `\\12\catcode `\$12\catcode `\&12\catcode `\#12\catcode `\^12\catcode `\_12\catcode `\%12\relax}%
\providecommand \@@startlink[1]{}%
\providecommand \@@endlink[0]{}%
\providecommand \url  [0]{\begingroup\@sanitize@url \@url }%
\providecommand \@url [1]{\endgroup\@href {#1}{\urlprefix }}%
\providecommand \urlprefix  [0]{URL }%
\providecommand \Eprint [0]{\href }%
\providecommand \doibase [0]{http://dx.doi.org/}%
\providecommand \selectlanguage [0]{\@gobble}%
\providecommand \bibinfo  [0]{\@secondoftwo}%
\providecommand \bibfield  [0]{\@secondoftwo}%
\providecommand \translation [1]{[#1]}%
\providecommand \BibitemOpen [0]{}%
\providecommand \bibitemStop [0]{}%
\providecommand \bibitemNoStop [0]{.\EOS\space}%
\providecommand \EOS [0]{\spacefactor3000\relax}%
\providecommand \BibitemShut  [1]{\csname bibitem#1\endcsname}%
\let\auto@bib@innerbib\@empty
\bibitem [{\citenamefont {Nagaosa}\ and\ \citenamefont {Heusler}(1999)}]{nagaosa}%
  \BibitemOpen
  \bibfield  {author} {\bibinfo {author} {\bibfnamefont {N.}~\bibnamefont {Nagaosa}}\ and\ \bibinfo {author} {\bibfnamefont {S.}~\bibnamefont {Heusler}},\ }\href {https://books.google.it/books?id=C9uAXYIlFhMC} {\emph {\bibinfo {title} {Quantum Field Theory in Condensed Matter Physics}}},\ Texts and monographs in physics\ (\bibinfo  {publisher} {Springer, New York, NY},\ \bibinfo {year} {1999})\BibitemShut {NoStop}%
\bibitem [{\citenamefont {Anderson}(1958)}]{anderson_pr58}%
  \BibitemOpen
  \bibfield  {author} {\bibinfo {author} {\bibfnamefont {P.~W.}\ \bibnamefont {Anderson}},\ }\href {\doibase 10.1103/PhysRev.112.1900} {\bibfield  {journal} {\bibinfo  {journal} {Phys. Rev.}\ }\textbf {\bibinfo {volume} {112}},\ \bibinfo {pages} {1900} (\bibinfo {year} {1958})}\BibitemShut {NoStop}%
\bibitem [{\citenamefont {Keimer}\ \emph {et~al.}(2015)\citenamefont {Keimer}, \citenamefont {Kivelson}, \citenamefont {Norman}, \citenamefont {Uchida},\ and\ \citenamefont {Zaanen}}]{keimer_review15}%
  \BibitemOpen
  \bibfield  {author} {\bibinfo {author} {\bibfnamefont {B.}~\bibnamefont {Keimer}}, \bibinfo {author} {\bibfnamefont {S.~A.}\ \bibnamefont {Kivelson}}, \bibinfo {author} {\bibfnamefont {M.~R.}\ \bibnamefont {Norman}}, \bibinfo {author} {\bibfnamefont {S.}~\bibnamefont {Uchida}}, \ and\ \bibinfo {author} {\bibfnamefont {J.}~\bibnamefont {Zaanen}},\ }\href {\doibase 10.1038/nature14165} {\bibfield  {journal} {\bibinfo  {journal} {Nature}\ }\textbf {\bibinfo {volume} {518}},\ \bibinfo {pages} {179} (\bibinfo {year} {2015})}\BibitemShut {NoStop}%
\bibitem [{\citenamefont {Homes}\ \emph {et~al.}(1993)\citenamefont {Homes}, \citenamefont {Timusk}, \citenamefont {Liang}, \citenamefont {Bonn},\ and\ \citenamefont {Hardy}}]{homes_prl93}%
  \BibitemOpen
  \bibfield  {author} {\bibinfo {author} {\bibfnamefont {C.~C.}\ \bibnamefont {Homes}}, \bibinfo {author} {\bibfnamefont {T.}~\bibnamefont {Timusk}}, \bibinfo {author} {\bibfnamefont {R.}~\bibnamefont {Liang}}, \bibinfo {author} {\bibfnamefont {D.~A.}\ \bibnamefont {Bonn}}, \ and\ \bibinfo {author} {\bibfnamefont {W.~N.}\ \bibnamefont {Hardy}},\ }\href {\doibase 10.1103/PhysRevLett.71.1645} {\bibfield  {journal} {\bibinfo  {journal} {Phys. Rev. Lett.}\ }\textbf {\bibinfo {volume} {71}},\ \bibinfo {pages} {1645} (\bibinfo {year} {1993})}\BibitemShut {NoStop}%
\bibitem [{\citenamefont {van~der Marel}\ and\ \citenamefont {Tsvetkov}(1996)}]{vandermarel96}%
  \BibitemOpen
  \bibfield  {author} {\bibinfo {author} {\bibfnamefont {D.}~\bibnamefont {van~der Marel}}\ and\ \bibinfo {author} {\bibfnamefont {A.}~\bibnamefont {Tsvetkov}},\ }\href {\doibase https://doi.org/10.1007/BF02548125} {\bibfield  {journal} {\bibinfo  {journal} {Czech. J. of Phys.}\ }\textbf {\bibinfo {volume} {46}},\ \bibinfo {pages} {3165} (\bibinfo {year} {1996})}\BibitemShut {NoStop}%
\bibitem [{\citenamefont {Shibata}\ and\ \citenamefont {Yamada}(1998)}]{yamada_prl98}%
  \BibitemOpen
  \bibfield  {author} {\bibinfo {author} {\bibfnamefont {H.}~\bibnamefont {Shibata}}\ and\ \bibinfo {author} {\bibfnamefont {T.}~\bibnamefont {Yamada}},\ }\href {\doibase 10.1103/PhysRevLett.81.3519} {\bibfield  {journal} {\bibinfo  {journal} {Phys. Rev. Lett.}\ }\textbf {\bibinfo {volume} {81}},\ \bibinfo {pages} {3519} (\bibinfo {year} {1998})}\BibitemShut {NoStop}%
\bibitem [{\citenamefont {Gr\"uninger}\ \emph {et~al.}(2000)\citenamefont {Gr\"uninger}, \citenamefont {van~der Marel}, \citenamefont {Tsvetkov},\ and\ \citenamefont {Erb}}]{erb_prl00}%
  \BibitemOpen
  \bibfield  {author} {\bibinfo {author} {\bibfnamefont {M.}~\bibnamefont {Gr\"uninger}}, \bibinfo {author} {\bibfnamefont {D.}~\bibnamefont {van~der Marel}}, \bibinfo {author} {\bibfnamefont {A.~A.}\ \bibnamefont {Tsvetkov}}, \ and\ \bibinfo {author} {\bibfnamefont {A.}~\bibnamefont {Erb}},\ }\href {\doibase 10.1103/PhysRevLett.84.1575} {\bibfield  {journal} {\bibinfo  {journal} {Phys. Rev. Lett.}\ }\textbf {\bibinfo {volume} {84}},\ \bibinfo {pages} {1575} (\bibinfo {year} {2000})}\BibitemShut {NoStop}%
\bibitem [{\citenamefont {van~der Marel}\ and\ \citenamefont {Tsvetkov}(2001)}]{vandermarel_prb01}%
  \BibitemOpen
  \bibfield  {author} {\bibinfo {author} {\bibfnamefont {D.}~\bibnamefont {van~der Marel}}\ and\ \bibinfo {author} {\bibfnamefont {A.~A.}\ \bibnamefont {Tsvetkov}},\ }\href {\doibase 10.1103/PhysRevB.64.024530} {\bibfield  {journal} {\bibinfo  {journal} {Phys. Rev. B}\ }\textbf {\bibinfo {volume} {64}},\ \bibinfo {pages} {024530} (\bibinfo {year} {2001})}\BibitemShut {NoStop}%
\bibitem [{\citenamefont {Dubroka}\ \emph {et~al.}(2011)\citenamefont {Dubroka}, \citenamefont {R\"ossle}, \citenamefont {Kim}, \citenamefont {Malik}, \citenamefont {Munzar}, \citenamefont {Basov}, \citenamefont {Schafgans}, \citenamefont {Moon}, \citenamefont {Lin}, \citenamefont {Haug}, \citenamefont {Hinkov}, \citenamefont {Keimer}, \citenamefont {Wolf}, \citenamefont {Storey}, \citenamefont {Tallon},\ and\ \citenamefont {Bernhard}}]{bernhard_prl11}%
  \BibitemOpen
  \bibfield  {author} {\bibinfo {author} {\bibfnamefont {A.}~\bibnamefont {Dubroka}}, \bibinfo {author} {\bibfnamefont {M.}~\bibnamefont {R\"ossle}}, \bibinfo {author} {\bibfnamefont {K.~W.}\ \bibnamefont {Kim}}, \bibinfo {author} {\bibfnamefont {V.~K.}\ \bibnamefont {Malik}}, \bibinfo {author} {\bibfnamefont {D.}~\bibnamefont {Munzar}}, \bibinfo {author} {\bibfnamefont {D.~N.}\ \bibnamefont {Basov}}, \bibinfo {author} {\bibfnamefont {A.~A.}\ \bibnamefont {Schafgans}}, \bibinfo {author} {\bibfnamefont {S.~J.}\ \bibnamefont {Moon}}, \bibinfo {author} {\bibfnamefont {C.~T.}\ \bibnamefont {Lin}}, \bibinfo {author} {\bibfnamefont {D.}~\bibnamefont {Haug}}, \bibinfo {author} {\bibfnamefont {V.}~\bibnamefont {Hinkov}}, \bibinfo {author} {\bibfnamefont {B.}~\bibnamefont {Keimer}}, \bibinfo {author} {\bibfnamefont {T.}~\bibnamefont {Wolf}}, \bibinfo {author} {\bibfnamefont {J.~G.}\ \bibnamefont {Storey}}, \bibinfo {author} {\bibfnamefont {J.~L.}\ \bibnamefont {Tallon}}, \ and\ \bibinfo {author} {\bibfnamefont
  {C.}~\bibnamefont {Bernhard}},\ }\href {\doibase 10.1103/PhysRevLett.106.047006} {\bibfield  {journal} {\bibinfo  {journal} {Phys. Rev. Lett.}\ }\textbf {\bibinfo {volume} {106}},\ \bibinfo {pages} {047006} (\bibinfo {year} {2011})}\BibitemShut {NoStop}%
\bibitem [{\citenamefont {Stinson}\ \emph {et~al.}(2014)\citenamefont {Stinson}, \citenamefont {Wu}, \citenamefont {Jiang}, \citenamefont {Fei}, \citenamefont {Rodin}, \citenamefont {Chapler}, \citenamefont {McLeod}, \citenamefont {Castro~Neto}, \citenamefont {Lee}, \citenamefont {Fogler},\ and\ \citenamefont {Basov}}]{basov_prb14}%
  \BibitemOpen
  \bibfield  {author} {\bibinfo {author} {\bibfnamefont {H.~T.}\ \bibnamefont {Stinson}}, \bibinfo {author} {\bibfnamefont {J.~S.}\ \bibnamefont {Wu}}, \bibinfo {author} {\bibfnamefont {B.~Y.}\ \bibnamefont {Jiang}}, \bibinfo {author} {\bibfnamefont {Z.}~\bibnamefont {Fei}}, \bibinfo {author} {\bibfnamefont {A.~S.}\ \bibnamefont {Rodin}}, \bibinfo {author} {\bibfnamefont {B.~C.}\ \bibnamefont {Chapler}}, \bibinfo {author} {\bibfnamefont {A.~S.}\ \bibnamefont {McLeod}}, \bibinfo {author} {\bibfnamefont {A.}~\bibnamefont {Castro~Neto}}, \bibinfo {author} {\bibfnamefont {Y.~S.}\ \bibnamefont {Lee}}, \bibinfo {author} {\bibfnamefont {M.~M.}\ \bibnamefont {Fogler}}, \ and\ \bibinfo {author} {\bibfnamefont {D.~N.}\ \bibnamefont {Basov}},\ }\href {\doibase 10.1103/PhysRevB.90.014502} {\bibfield  {journal} {\bibinfo  {journal} {Phys. Rev. B}\ }\textbf {\bibinfo {volume} {90}},\ \bibinfo {pages} {014502} (\bibinfo {year} {2014})}\BibitemShut {NoStop}%
\bibitem [{\citenamefont {Lu}\ \emph {et~al.}(2020)\citenamefont {Lu}, \citenamefont {Bollinger}, \citenamefont {He}, \citenamefont {Sundling}, \citenamefont {Bozovic},\ and\ \citenamefont {Gozar}}]{gozar_nqm21}%
  \BibitemOpen
  \bibfield  {author} {\bibinfo {author} {\bibfnamefont {Q.}~\bibnamefont {Lu}}, \bibinfo {author} {\bibfnamefont {A.~T.}\ \bibnamefont {Bollinger}}, \bibinfo {author} {\bibfnamefont {X.}~\bibnamefont {He}}, \bibinfo {author} {\bibfnamefont {R.}~\bibnamefont {Sundling}}, \bibinfo {author} {\bibfnamefont {I.}~\bibnamefont {Bozovic}}, \ and\ \bibinfo {author} {\bibfnamefont {A.}~\bibnamefont {Gozar}},\ }\href {\doibase 10.1038/s41535-020-00272-8} {\bibfield  {journal} {\bibinfo  {journal} {npj Quantum Materials}\ }\textbf {\bibinfo {volume} {5}},\ \bibinfo {pages} {69} (\bibinfo {year} {2020})}\BibitemShut {NoStop}%
\bibitem [{\citenamefont {Savel'ev}\ \emph {et~al.}(2010)\citenamefont {Savel'ev}, \citenamefont {Yampol'skii}, \citenamefont {Rakhmanov},\ and\ \citenamefont {Nori}}]{nori_review10}%
  \BibitemOpen
  \bibfield  {author} {\bibinfo {author} {\bibfnamefont {S.}~\bibnamefont {Savel'ev}}, \bibinfo {author} {\bibfnamefont {V.~A.}\ \bibnamefont {Yampol'skii}}, \bibinfo {author} {\bibfnamefont {A.~L.}\ \bibnamefont {Rakhmanov}}, \ and\ \bibinfo {author} {\bibfnamefont {F.}~\bibnamefont {Nori}},\ }\href {\doibase 10.1088/0034-4885/73/2/026501} {\bibfield  {journal} {\bibinfo  {journal} {Reports on Progress in Physics}\ }\textbf {\bibinfo {volume} {73}},\ \bibinfo {pages} {026501} (\bibinfo {year} {2010})}\BibitemShut {NoStop}%
\bibitem [{\citenamefont {Hu}\ \emph {et~al.}(2014)\citenamefont {Hu}, \citenamefont {Kaiser}, \citenamefont {Nicoletti}, \citenamefont {Hunt}, \citenamefont {Gierz}, \citenamefont {Hoffmann}, \citenamefont {Le~Tacon}, \citenamefont {Loew}, \citenamefont {Keimer},\ and\ \citenamefont {Cavalleri}}]{cavalleri_nmat14}%
  \BibitemOpen
  \bibfield  {author} {\bibinfo {author} {\bibfnamefont {H.}~\bibnamefont {Hu}}, \bibinfo {author} {\bibfnamefont {S.}~\bibnamefont {Kaiser}}, \bibinfo {author} {\bibfnamefont {D.}~\bibnamefont {Nicoletti}}, \bibinfo {author} {\bibfnamefont {C.~R.}\ \bibnamefont {Hunt}}, \bibinfo {author} {\bibfnamefont {I.}~\bibnamefont {Gierz}}, \bibinfo {author} {\bibfnamefont {M.~C.}\ \bibnamefont {Hoffmann}}, \bibinfo {author} {\bibfnamefont {M.}~\bibnamefont {Le~Tacon}}, \bibinfo {author} {\bibfnamefont {T.}~\bibnamefont {Loew}}, \bibinfo {author} {\bibfnamefont {B.}~\bibnamefont {Keimer}}, \ and\ \bibinfo {author} {\bibfnamefont {A.}~\bibnamefont {Cavalleri}},\ }\href {\doibase 10.1038/nmat3963} {\bibfield  {journal} {\bibinfo  {journal} {Nature Materials}\ }\textbf {\bibinfo {volume} {13}},\ \bibinfo {pages} {705} (\bibinfo {year} {2014})}\BibitemShut {NoStop}%
\bibitem [{\citenamefont {Rajasekaran}\ \emph {et~al.}(2018)\citenamefont {Rajasekaran}, \citenamefont {Okamoto}, \citenamefont {Mathey}, \citenamefont {Fechner}, \citenamefont {Thampy}, \citenamefont {Gu},\ and\ \citenamefont {Cavalleri}}]{cavalleri_science18}%
  \BibitemOpen
  \bibfield  {author} {\bibinfo {author} {\bibfnamefont {S.}~\bibnamefont {Rajasekaran}}, \bibinfo {author} {\bibfnamefont {J.}~\bibnamefont {Okamoto}}, \bibinfo {author} {\bibfnamefont {L.}~\bibnamefont {Mathey}}, \bibinfo {author} {\bibfnamefont {M.}~\bibnamefont {Fechner}}, \bibinfo {author} {\bibfnamefont {V.}~\bibnamefont {Thampy}}, \bibinfo {author} {\bibfnamefont {G.~D.}\ \bibnamefont {Gu}}, \ and\ \bibinfo {author} {\bibfnamefont {A.}~\bibnamefont {Cavalleri}},\ }\href {\doibase 10.1126/science.aan3438} {\bibfield  {journal} {\bibinfo  {journal} {Science}\ }\textbf {\bibinfo {volume} {359}},\ \bibinfo {pages} {575} (\bibinfo {year} {2018})}\BibitemShut {NoStop}%
\bibitem [{\citenamefont {Zhang}\ \emph {et~al.}(2020)\citenamefont {Zhang}, \citenamefont {Wang}, \citenamefont {Xiang}, \citenamefont {Yao}, \citenamefont {Liu}, \citenamefont {Shi}, \citenamefont {Lin}, \citenamefont {Dong}, \citenamefont {Wu},\ and\ \citenamefont {Wang}}]{wangNL_prx20}%
  \BibitemOpen
  \bibfield  {author} {\bibinfo {author} {\bibfnamefont {S.~J.}\ \bibnamefont {Zhang}}, \bibinfo {author} {\bibfnamefont {Z.~X.}\ \bibnamefont {Wang}}, \bibinfo {author} {\bibfnamefont {H.}~\bibnamefont {Xiang}}, \bibinfo {author} {\bibfnamefont {X.}~\bibnamefont {Yao}}, \bibinfo {author} {\bibfnamefont {Q.~M.}\ \bibnamefont {Liu}}, \bibinfo {author} {\bibfnamefont {L.~Y.}\ \bibnamefont {Shi}}, \bibinfo {author} {\bibfnamefont {T.}~\bibnamefont {Lin}}, \bibinfo {author} {\bibfnamefont {T.}~\bibnamefont {Dong}}, \bibinfo {author} {\bibfnamefont {D.}~\bibnamefont {Wu}}, \ and\ \bibinfo {author} {\bibfnamefont {N.~L.}\ \bibnamefont {Wang}},\ }\href {\doibase 10.1103/PhysRevX.10.011056} {\bibfield  {journal} {\bibinfo  {journal} {Phys. Rev. X}\ }\textbf {\bibinfo {volume} {10}},\ \bibinfo {pages} {011056} (\bibinfo {year} {2020})}\BibitemShut {NoStop}%
\bibitem [{\citenamefont {Katsumi}\ \emph {et~al.}(2023)\citenamefont {Katsumi}, \citenamefont {Nishida}, \citenamefont {Kaiser}, \citenamefont {Miyasaka}, \citenamefont {Tajima},\ and\ \citenamefont {Shimano}}]{shimano_prb23}%
  \BibitemOpen
  \bibfield  {author} {\bibinfo {author} {\bibfnamefont {K.}~\bibnamefont {Katsumi}}, \bibinfo {author} {\bibfnamefont {M.}~\bibnamefont {Nishida}}, \bibinfo {author} {\bibfnamefont {S.}~\bibnamefont {Kaiser}}, \bibinfo {author} {\bibfnamefont {S.}~\bibnamefont {Miyasaka}}, \bibinfo {author} {\bibfnamefont {S.}~\bibnamefont {Tajima}}, \ and\ \bibinfo {author} {\bibfnamefont {R.}~\bibnamefont {Shimano}},\ }\href {\doibase 10.1103/PhysRevB.107.214506} {\bibfield  {journal} {\bibinfo  {journal} {Phys. Rev. B}\ }\textbf {\bibinfo {volume} {107}},\ \bibinfo {pages} {214506} (\bibinfo {year} {2023})}\BibitemShut {NoStop}%
\bibitem [{\citenamefont {Kaj}\ \emph {et~al.}(2023)\citenamefont {Kaj}, \citenamefont {Cremin}, \citenamefont {Hammock}, \citenamefont {Schalch}, \citenamefont {Basov},\ and\ \citenamefont {Averitt}}]{averitt_prb23}%
  \BibitemOpen
  \bibfield  {author} {\bibinfo {author} {\bibfnamefont {K.}~\bibnamefont {Kaj}}, \bibinfo {author} {\bibfnamefont {K.~A.}\ \bibnamefont {Cremin}}, \bibinfo {author} {\bibfnamefont {I.}~\bibnamefont {Hammock}}, \bibinfo {author} {\bibfnamefont {J.}~\bibnamefont {Schalch}}, \bibinfo {author} {\bibfnamefont {D.~N.}\ \bibnamefont {Basov}}, \ and\ \bibinfo {author} {\bibfnamefont {R.~D.}\ \bibnamefont {Averitt}},\ }\href {\doibase 10.1103/PhysRevB.107.L140504} {\bibfield  {journal} {\bibinfo  {journal} {Phys. Rev. B}\ }\textbf {\bibinfo {volume} {107}},\ \bibinfo {pages} {L140504} (\bibinfo {year} {2023})}\BibitemShut {NoStop}%
\bibitem [{\citenamefont {Gabriele}\ \emph {et~al.}(2022)\citenamefont {Gabriele}, \citenamefont {Castellani},\ and\ \citenamefont {Benfatto}}]{gabriele_prr22}%
  \BibitemOpen
  \bibfield  {author} {\bibinfo {author} {\bibfnamefont {F.}~\bibnamefont {Gabriele}}, \bibinfo {author} {\bibfnamefont {C.}~\bibnamefont {Castellani}}, \ and\ \bibinfo {author} {\bibfnamefont {L.}~\bibnamefont {Benfatto}},\ }\href {\doibase 10.1103/PhysRevResearch.4.023112} {\bibfield  {journal} {\bibinfo  {journal} {Phys. Rev. Res.}\ }\textbf {\bibinfo {volume} {4}},\ \bibinfo {pages} {023112} (\bibinfo {year} {2022})}\BibitemShut {NoStop}%
\bibitem [{\citenamefont {Sellati}\ \emph {et~al.}(2023)\citenamefont {Sellati}, \citenamefont {Gabriele}, \citenamefont {Castellani},\ and\ \citenamefont {Benfatto}}]{sellati_prb23}%
  \BibitemOpen
  \bibfield  {author} {\bibinfo {author} {\bibfnamefont {N.}~\bibnamefont {Sellati}}, \bibinfo {author} {\bibfnamefont {F.}~\bibnamefont {Gabriele}}, \bibinfo {author} {\bibfnamefont {C.}~\bibnamefont {Castellani}}, \ and\ \bibinfo {author} {\bibfnamefont {L.}~\bibnamefont {Benfatto}},\ }\href {\doibase 10.1103/PhysRevB.108.014503} {\bibfield  {journal} {\bibinfo  {journal} {Phys. Rev. B}\ }\textbf {\bibinfo {volume} {108}},\ \bibinfo {pages} {014503} (\bibinfo {year} {2023})}\BibitemShut {NoStop}%
\bibitem [{\citenamefont {Gabriele}\ \emph {et~al.}(2024)\citenamefont {Gabriele}, \citenamefont {Senese}, \citenamefont {Castellani},\ and\ \citenamefont {Benfatto}}]{gabriele_prb24}%
  \BibitemOpen
  \bibfield  {author} {\bibinfo {author} {\bibfnamefont {F.}~\bibnamefont {Gabriele}}, \bibinfo {author} {\bibfnamefont {R.}~\bibnamefont {Senese}}, \bibinfo {author} {\bibfnamefont {C.}~\bibnamefont {Castellani}}, \ and\ \bibinfo {author} {\bibfnamefont {L.}~\bibnamefont {Benfatto}},\ }\href {\doibase 10.1103/PhysRevB.109.045137} {\bibfield  {journal} {\bibinfo  {journal} {Phys. Rev. B}\ }\textbf {\bibinfo {volume} {109}},\ \bibinfo {pages} {045137} (\bibinfo {year} {2024})}\BibitemShut {NoStop}%
\bibitem [{\citenamefont {Bulaevskii}\ \emph {et~al.}(1994)\citenamefont {Bulaevskii}, \citenamefont {Zamora}, \citenamefont {Baeriswyl}, \citenamefont {Beck},\ and\ \citenamefont {Clem}}]{bulaevskii_prb94}%
  \BibitemOpen
  \bibfield  {author} {\bibinfo {author} {\bibfnamefont {L.~N.}\ \bibnamefont {Bulaevskii}}, \bibinfo {author} {\bibfnamefont {M.}~\bibnamefont {Zamora}}, \bibinfo {author} {\bibfnamefont {D.}~\bibnamefont {Baeriswyl}}, \bibinfo {author} {\bibfnamefont {H.}~\bibnamefont {Beck}}, \ and\ \bibinfo {author} {\bibfnamefont {J.~R.}\ \bibnamefont {Clem}},\ }\href {\doibase 10.1103/PhysRevB.50.12831} {\bibfield  {journal} {\bibinfo  {journal} {Phys. Rev. B}\ }\textbf {\bibinfo {volume} {50}},\ \bibinfo {pages} {12831} (\bibinfo {year} {1994})}\BibitemShut {NoStop}%
\bibitem [{\citenamefont {Helm}\ and\ \citenamefont {Bulaevskii}(2002)}]{bulaevskii_prb02}%
  \BibitemOpen
  \bibfield  {author} {\bibinfo {author} {\bibfnamefont {C.}~\bibnamefont {Helm}}\ and\ \bibinfo {author} {\bibfnamefont {L.~N.}\ \bibnamefont {Bulaevskii}},\ }\href {\doibase 10.1103/PhysRevB.66.094514} {\bibfield  {journal} {\bibinfo  {journal} {Phys. Rev. B}\ }\textbf {\bibinfo {volume} {66}},\ \bibinfo {pages} {094514} (\bibinfo {year} {2002})}\BibitemShut {NoStop}%
\bibitem [{\citenamefont {Laplace}\ and\ \citenamefont {Cavalleri}(2016)}]{cavalleri_review}%
  \BibitemOpen
  \bibfield  {author} {\bibinfo {author} {\bibfnamefont {Y.}~\bibnamefont {Laplace}}\ and\ \bibinfo {author} {\bibfnamefont {A.}~\bibnamefont {Cavalleri}},\ }\href {\doibase 10.1080/23746149.2016.1212671} {\bibfield  {journal} {\bibinfo  {journal} {Advances in Physics: X}\ }\textbf {\bibinfo {volume} {1}},\ \bibinfo {pages} {387} (\bibinfo {year} {2016})}\BibitemShut {NoStop}%
\bibitem [{\citenamefont {Salvador}\ \emph {et~al.}(2024)\citenamefont {Salvador}, \citenamefont {Dolgirev}, \citenamefont {Michael}, \citenamefont {Liu}, \citenamefont {Pavicevic}, \citenamefont {Fechner}, \citenamefont {Cavalleri},\ and\ \citenamefont {Demler}}]{demler_cm24}%
  \BibitemOpen
  \bibfield  {author} {\bibinfo {author} {\bibfnamefont {A.~G.}\ \bibnamefont {Salvador}}, \bibinfo {author} {\bibfnamefont {P.~E.}\ \bibnamefont {Dolgirev}}, \bibinfo {author} {\bibfnamefont {M.~H.}\ \bibnamefont {Michael}}, \bibinfo {author} {\bibfnamefont {A.}~\bibnamefont {Liu}}, \bibinfo {author} {\bibfnamefont {D.}~\bibnamefont {Pavicevic}}, \bibinfo {author} {\bibfnamefont {M.}~\bibnamefont {Fechner}}, \bibinfo {author} {\bibfnamefont {A.}~\bibnamefont {Cavalleri}}, \ and\ \bibinfo {author} {\bibfnamefont {E.}~\bibnamefont {Demler}},\ }\href@noop {} {\enquote {\bibinfo {title} {Principles of 2d terahertz spectroscopy of collective excitations: the case of josephson plasmons in layered superconductors},}\ } (\bibinfo {year} {2024}),\ \Eprint {http://arxiv.org/abs/2401.05503} {arXiv:2401.05503 [cond-mat.supr-con]} \BibitemShut {NoStop}%
\bibitem [{\citenamefont {Hepting}\ \emph {et~al.}(2018)\citenamefont {Hepting}, \citenamefont {Chaix}, \citenamefont {Huang}, \citenamefont {Fumagalli}, \citenamefont {Peng}, \citenamefont {Moritz}, \citenamefont {Kummer}, \citenamefont {Brookes}, \citenamefont {Lee}, \citenamefont {Hashimoto}, \citenamefont {Sarkar}, \citenamefont {He}, \citenamefont {Rotundu}, \citenamefont {Lee}, \citenamefont {Greene}, \citenamefont {Braicovich}, \citenamefont {Ghiringhelli}, \citenamefont {Shen}, \citenamefont {Devereaux},\ and\ \citenamefont {Lee}}]{lee_rixs_nature18}%
  \BibitemOpen
  \bibfield  {author} {\bibinfo {author} {\bibfnamefont {M.}~\bibnamefont {Hepting}}, \bibinfo {author} {\bibfnamefont {L.}~\bibnamefont {Chaix}}, \bibinfo {author} {\bibfnamefont {E.~W.}\ \bibnamefont {Huang}}, \bibinfo {author} {\bibfnamefont {R.}~\bibnamefont {Fumagalli}}, \bibinfo {author} {\bibfnamefont {Y.~Y.}\ \bibnamefont {Peng}}, \bibinfo {author} {\bibfnamefont {B.}~\bibnamefont {Moritz}}, \bibinfo {author} {\bibfnamefont {K.}~\bibnamefont {Kummer}}, \bibinfo {author} {\bibfnamefont {N.~B.}\ \bibnamefont {Brookes}}, \bibinfo {author} {\bibfnamefont {W.~C.}\ \bibnamefont {Lee}}, \bibinfo {author} {\bibfnamefont {M.}~\bibnamefont {Hashimoto}}, \bibinfo {author} {\bibfnamefont {T.}~\bibnamefont {Sarkar}}, \bibinfo {author} {\bibfnamefont {J.~F.}\ \bibnamefont {He}}, \bibinfo {author} {\bibfnamefont {C.~R.}\ \bibnamefont {Rotundu}}, \bibinfo {author} {\bibfnamefont {Y.~S.}\ \bibnamefont {Lee}}, \bibinfo {author} {\bibfnamefont {R.~L.}\ \bibnamefont {Greene}}, \bibinfo {author} {\bibfnamefont
  {L.}~\bibnamefont {Braicovich}}, \bibinfo {author} {\bibfnamefont {G.}~\bibnamefont {Ghiringhelli}}, \bibinfo {author} {\bibfnamefont {Z.~X.}\ \bibnamefont {Shen}}, \bibinfo {author} {\bibfnamefont {T.~P.}\ \bibnamefont {Devereaux}}, \ and\ \bibinfo {author} {\bibfnamefont {W.~S.}\ \bibnamefont {Lee}},\ }\href {\doibase 10.1038/s41586-018-0648-3} {\bibfield  {journal} {\bibinfo  {journal} {Nature}\ }\textbf {\bibinfo {volume} {563}},\ \bibinfo {pages} {374} (\bibinfo {year} {2018})}\BibitemShut {NoStop}%
\bibitem [{\citenamefont {Lin}\ \emph {et~al.}(2020)\citenamefont {Lin}, \citenamefont {Yuan}, \citenamefont {Jin}, \citenamefont {Yin}, \citenamefont {Li}, \citenamefont {Zhou}, \citenamefont {Lu}, \citenamefont {Dantz}, \citenamefont {Schmitt}, \citenamefont {Ding}, \citenamefont {Guo}, \citenamefont {Dean},\ and\ \citenamefont {Liu}}]{liu_rixs_npjqm20}%
  \BibitemOpen
  \bibfield  {author} {\bibinfo {author} {\bibfnamefont {J.}~\bibnamefont {Lin}}, \bibinfo {author} {\bibfnamefont {J.}~\bibnamefont {Yuan}}, \bibinfo {author} {\bibfnamefont {K.}~\bibnamefont {Jin}}, \bibinfo {author} {\bibfnamefont {Z.}~\bibnamefont {Yin}}, \bibinfo {author} {\bibfnamefont {G.}~\bibnamefont {Li}}, \bibinfo {author} {\bibfnamefont {K.-J.}\ \bibnamefont {Zhou}}, \bibinfo {author} {\bibfnamefont {X.}~\bibnamefont {Lu}}, \bibinfo {author} {\bibfnamefont {M.}~\bibnamefont {Dantz}}, \bibinfo {author} {\bibfnamefont {T.}~\bibnamefont {Schmitt}}, \bibinfo {author} {\bibfnamefont {H.}~\bibnamefont {Ding}}, \bibinfo {author} {\bibfnamefont {H.}~\bibnamefont {Guo}}, \bibinfo {author} {\bibfnamefont {M.~P.~M.}\ \bibnamefont {Dean}}, \ and\ \bibinfo {author} {\bibfnamefont {X.}~\bibnamefont {Liu}},\ }\href {\doibase 10.1038/s41535-019-0205-9} {\bibfield  {journal} {\bibinfo  {journal} {npj Quantum Materials}\ }\textbf {\bibinfo {volume} {5}},\ \bibinfo {pages} {4} (\bibinfo {year} {2020})}\BibitemShut
  {NoStop}%
\bibitem [{\citenamefont {Nag}\ \emph {et~al.}(2020)\citenamefont {Nag}, \citenamefont {Zhu}, \citenamefont {Bejas}, \citenamefont {Li}, \citenamefont {Robarts}, \citenamefont {Yamase}, \citenamefont {Petsch}, \citenamefont {Song}, \citenamefont {Eisaki}, \citenamefont {Walters}, \citenamefont {Garc\'{\i}a-Fern\'andez}, \citenamefont {Greco}, \citenamefont {Hayden},\ and\ \citenamefont {Zhou}}]{zhou_prl20}%
  \BibitemOpen
  \bibfield  {author} {\bibinfo {author} {\bibfnamefont {A.}~\bibnamefont {Nag}}, \bibinfo {author} {\bibfnamefont {M.}~\bibnamefont {Zhu}}, \bibinfo {author} {\bibfnamefont {M.}~\bibnamefont {Bejas}}, \bibinfo {author} {\bibfnamefont {J.}~\bibnamefont {Li}}, \bibinfo {author} {\bibfnamefont {H.~C.}\ \bibnamefont {Robarts}}, \bibinfo {author} {\bibfnamefont {H.}~\bibnamefont {Yamase}}, \bibinfo {author} {\bibfnamefont {A.~N.}\ \bibnamefont {Petsch}}, \bibinfo {author} {\bibfnamefont {D.}~\bibnamefont {Song}}, \bibinfo {author} {\bibfnamefont {H.}~\bibnamefont {Eisaki}}, \bibinfo {author} {\bibfnamefont {A.~C.}\ \bibnamefont {Walters}}, \bibinfo {author} {\bibfnamefont {M.}~\bibnamefont {Garc\'{\i}a-Fern\'andez}}, \bibinfo {author} {\bibfnamefont {A.}~\bibnamefont {Greco}}, \bibinfo {author} {\bibfnamefont {S.~M.}\ \bibnamefont {Hayden}}, \ and\ \bibinfo {author} {\bibfnamefont {K.-J.}\ \bibnamefont {Zhou}},\ }\href {\doibase 10.1103/PhysRevLett.125.257002} {\bibfield  {journal} {\bibinfo  {journal} {Phys.
  Rev. Lett.}\ }\textbf {\bibinfo {volume} {125}},\ \bibinfo {pages} {257002} (\bibinfo {year} {2020})}\BibitemShut {NoStop}%
\bibitem [{\citenamefont {Singh}\ \emph {et~al.}(2022)\citenamefont {Singh}, \citenamefont {Huang}, \citenamefont {Lane}, \citenamefont {Li}, \citenamefont {Okamoto}, \citenamefont {Komiya}, \citenamefont {Markiewicz}, \citenamefont {Bansil}, \citenamefont {Lee}, \citenamefont {Fujimori}, \citenamefont {Chen},\ and\ \citenamefont {Huang}}]{huang_rixs_prb22}%
  \BibitemOpen
  \bibfield  {author} {\bibinfo {author} {\bibfnamefont {A.}~\bibnamefont {Singh}}, \bibinfo {author} {\bibfnamefont {H.~Y.}\ \bibnamefont {Huang}}, \bibinfo {author} {\bibfnamefont {C.}~\bibnamefont {Lane}}, \bibinfo {author} {\bibfnamefont {J.~H.}\ \bibnamefont {Li}}, \bibinfo {author} {\bibfnamefont {J.}~\bibnamefont {Okamoto}}, \bibinfo {author} {\bibfnamefont {S.}~\bibnamefont {Komiya}}, \bibinfo {author} {\bibfnamefont {R.~S.}\ \bibnamefont {Markiewicz}}, \bibinfo {author} {\bibfnamefont {A.}~\bibnamefont {Bansil}}, \bibinfo {author} {\bibfnamefont {T.~K.}\ \bibnamefont {Lee}}, \bibinfo {author} {\bibfnamefont {A.}~\bibnamefont {Fujimori}}, \bibinfo {author} {\bibfnamefont {C.~T.}\ \bibnamefont {Chen}}, \ and\ \bibinfo {author} {\bibfnamefont {D.~J.}\ \bibnamefont {Huang}},\ }\href {\doibase 10.1103/PhysRevB.105.235105} {\bibfield  {journal} {\bibinfo  {journal} {Phys. Rev. B}\ }\textbf {\bibinfo {volume} {105}},\ \bibinfo {pages} {235105} (\bibinfo {year} {2022})}\BibitemShut {NoStop}%
\bibitem [{\citenamefont {Bejas}\ \emph {et~al.}(2023)\citenamefont {Bejas}, \citenamefont {Zimmermann}, \citenamefont {Betto}, \citenamefont {Boyko}, \citenamefont {Green}, \citenamefont {Loew}, \citenamefont {Brookes}, \citenamefont {Cristiani}, \citenamefont {Logvenov}, \citenamefont {Minola}, \citenamefont {Keimer}, \citenamefont {Yamase}, \citenamefont {Greco},\ and\ \citenamefont {Hepting}}]{hepting_rixs_cm23}%
  \BibitemOpen
  \bibfield  {author} {\bibinfo {author} {\bibfnamefont {M.}~\bibnamefont {Bejas}}, \bibinfo {author} {\bibfnamefont {V.}~\bibnamefont {Zimmermann}}, \bibinfo {author} {\bibfnamefont {D.}~\bibnamefont {Betto}}, \bibinfo {author} {\bibfnamefont {T.~D.}\ \bibnamefont {Boyko}}, \bibinfo {author} {\bibfnamefont {R.~J.}\ \bibnamefont {Green}}, \bibinfo {author} {\bibfnamefont {T.}~\bibnamefont {Loew}}, \bibinfo {author} {\bibfnamefont {N.~B.}\ \bibnamefont {Brookes}}, \bibinfo {author} {\bibfnamefont {G.}~\bibnamefont {Cristiani}}, \bibinfo {author} {\bibfnamefont {G.}~\bibnamefont {Logvenov}}, \bibinfo {author} {\bibfnamefont {M.}~\bibnamefont {Minola}}, \bibinfo {author} {\bibfnamefont {B.}~\bibnamefont {Keimer}}, \bibinfo {author} {\bibfnamefont {H.}~\bibnamefont {Yamase}}, \bibinfo {author} {\bibfnamefont {A.}~\bibnamefont {Greco}}, \ and\ \bibinfo {author} {\bibfnamefont {M.}~\bibnamefont {Hepting}},\ }\href@noop {} {\enquote {\bibinfo {title} {Plasmon dispersion in bilayer cuprate superconductors},}\ }
  (\bibinfo {year} {2023}),\ \Eprint {http://arxiv.org/abs/2311.01413} {arXiv:2311.01413 [cond-mat.supr-con]} \BibitemShut {NoStop}%
\bibitem [{\citenamefont {Mitrano}\ \emph {et~al.}(2018)\citenamefont {Mitrano}, \citenamefont {Husain}, \citenamefont {Vig}, \citenamefont {Kogar}, \citenamefont {Rak}, \citenamefont {Rubeck}, \citenamefont {Schmalian}, \citenamefont {Uchoa}, \citenamefont {Schneeloch}, \citenamefont {Zhong}, \citenamefont {Gu},\ and\ \citenamefont {Abbamonte}}]{mitrano_pnas18}%
  \BibitemOpen
  \bibfield  {author} {\bibinfo {author} {\bibfnamefont {M.}~\bibnamefont {Mitrano}}, \bibinfo {author} {\bibfnamefont {A.~A.}\ \bibnamefont {Husain}}, \bibinfo {author} {\bibfnamefont {S.}~\bibnamefont {Vig}}, \bibinfo {author} {\bibfnamefont {A.}~\bibnamefont {Kogar}}, \bibinfo {author} {\bibfnamefont {M.~S.}\ \bibnamefont {Rak}}, \bibinfo {author} {\bibfnamefont {S.~I.}\ \bibnamefont {Rubeck}}, \bibinfo {author} {\bibfnamefont {J.}~\bibnamefont {Schmalian}}, \bibinfo {author} {\bibfnamefont {B.}~\bibnamefont {Uchoa}}, \bibinfo {author} {\bibfnamefont {J.}~\bibnamefont {Schneeloch}}, \bibinfo {author} {\bibfnamefont {R.}~\bibnamefont {Zhong}}, \bibinfo {author} {\bibfnamefont {G.~D.}\ \bibnamefont {Gu}}, \ and\ \bibinfo {author} {\bibfnamefont {P.}~\bibnamefont {Abbamonte}},\ }\href {\doibase 10.1073/pnas.1721495115} {\bibfield  {journal} {\bibinfo  {journal} {Proceedings of the National Academy of Sciences}\ }\textbf {\bibinfo {volume} {115}},\ \bibinfo {pages} {5392} (\bibinfo {year} {2018})}\BibitemShut
  {NoStop}%
\bibitem [{\citenamefont {Husain}\ \emph {et~al.}(2019)\citenamefont {Husain}, \citenamefont {Mitrano}, \citenamefont {Rak}, \citenamefont {Rubeck}, \citenamefont {Uchoa}, \citenamefont {March}, \citenamefont {Dwyer}, \citenamefont {Schneeloch}, \citenamefont {Zhong}, \citenamefont {Gu},\ and\ \citenamefont {Abbamonte}}]{mitrano_prx19}%
  \BibitemOpen
  \bibfield  {author} {\bibinfo {author} {\bibfnamefont {A.~A.}\ \bibnamefont {Husain}}, \bibinfo {author} {\bibfnamefont {M.}~\bibnamefont {Mitrano}}, \bibinfo {author} {\bibfnamefont {M.~S.}\ \bibnamefont {Rak}}, \bibinfo {author} {\bibfnamefont {S.}~\bibnamefont {Rubeck}}, \bibinfo {author} {\bibfnamefont {B.}~\bibnamefont {Uchoa}}, \bibinfo {author} {\bibfnamefont {K.}~\bibnamefont {March}}, \bibinfo {author} {\bibfnamefont {C.}~\bibnamefont {Dwyer}}, \bibinfo {author} {\bibfnamefont {J.}~\bibnamefont {Schneeloch}}, \bibinfo {author} {\bibfnamefont {R.}~\bibnamefont {Zhong}}, \bibinfo {author} {\bibfnamefont {G.~D.}\ \bibnamefont {Gu}}, \ and\ \bibinfo {author} {\bibfnamefont {P.}~\bibnamefont {Abbamonte}},\ }\href {\doibase 10.1103/PhysRevX.9.041062} {\bibfield  {journal} {\bibinfo  {journal} {Phys. Rev. X}\ }\textbf {\bibinfo {volume} {9}},\ \bibinfo {pages} {041062} (\bibinfo {year} {2019})}\BibitemShut {NoStop}%
\bibitem [{\citenamefont {Thornton}\ \emph {et~al.}(2023)\citenamefont {Thornton}, \citenamefont {Liarte}, \citenamefont {Abbamonte}, \citenamefont {Sethna},\ and\ \citenamefont {Chowdhury}}]{abbamonte_natcomm23}%
  \BibitemOpen
  \bibfield  {author} {\bibinfo {author} {\bibfnamefont {S.~J.}\ \bibnamefont {Thornton}}, \bibinfo {author} {\bibfnamefont {D.~B.}\ \bibnamefont {Liarte}}, \bibinfo {author} {\bibfnamefont {P.}~\bibnamefont {Abbamonte}}, \bibinfo {author} {\bibfnamefont {J.~P.}\ \bibnamefont {Sethna}}, \ and\ \bibinfo {author} {\bibfnamefont {D.}~\bibnamefont {Chowdhury}},\ }\href {\doibase 10.1038/s41467-023-39499-x} {\bibfield  {journal} {\bibinfo  {journal} {Nature Communications}\ }\textbf {\bibinfo {volume} {14}},\ \bibinfo {pages} {3919} (\bibinfo {year} {2023})}\BibitemShut {NoStop}%
\bibitem [{\citenamefont {Pimenov}\ \emph {et~al.}(2000{\natexlab{a}})\citenamefont {Pimenov}, \citenamefont {Pronin}, \citenamefont {Loidl}, \citenamefont {Kampf}, \citenamefont {Krasnosvobodtsev},\ and\ \citenamefont {Nozdrin}}]{pimenov_apl00}%
  \BibitemOpen
  \bibfield  {author} {\bibinfo {author} {\bibfnamefont {A.}~\bibnamefont {Pimenov}}, \bibinfo {author} {\bibfnamefont {A.~V.}\ \bibnamefont {Pronin}}, \bibinfo {author} {\bibfnamefont {A.}~\bibnamefont {Loidl}}, \bibinfo {author} {\bibfnamefont {A.~P.}\ \bibnamefont {Kampf}}, \bibinfo {author} {\bibfnamefont {S.~I.}\ \bibnamefont {Krasnosvobodtsev}}, \ and\ \bibinfo {author} {\bibfnamefont {V.~S.}\ \bibnamefont {Nozdrin}},\ }\href {\doibase 10.1063/1.126999} {\bibfield  {journal} {\bibinfo  {journal} {Applied Physics Letters}\ }\textbf {\bibinfo {volume} {77}},\ \bibinfo {pages} {429} (\bibinfo {year} {2000}{\natexlab{a}})}\BibitemShut {NoStop}%
\bibitem [{\citenamefont {Pimenov}\ \emph {et~al.}(2000{\natexlab{b}})\citenamefont {Pimenov}, \citenamefont {Pronin}, \citenamefont {Loidl}, \citenamefont {Michelucci}, \citenamefont {Kampf}, \citenamefont {Krasnosvobodtsev}, \citenamefont {Nozdrin},\ and\ \citenamefont {Rainer}}]{pimenov_prb00}%
  \BibitemOpen
  \bibfield  {author} {\bibinfo {author} {\bibfnamefont {A.}~\bibnamefont {Pimenov}}, \bibinfo {author} {\bibfnamefont {A.~V.}\ \bibnamefont {Pronin}}, \bibinfo {author} {\bibfnamefont {A.}~\bibnamefont {Loidl}}, \bibinfo {author} {\bibfnamefont {U.}~\bibnamefont {Michelucci}}, \bibinfo {author} {\bibfnamefont {A.~P.}\ \bibnamefont {Kampf}}, \bibinfo {author} {\bibfnamefont {S.~I.}\ \bibnamefont {Krasnosvobodtsev}}, \bibinfo {author} {\bibfnamefont {V.~S.}\ \bibnamefont {Nozdrin}}, \ and\ \bibinfo {author} {\bibfnamefont {D.}~\bibnamefont {Rainer}},\ }\href {\doibase 10.1103/PhysRevB.62.9822} {\bibfield  {journal} {\bibinfo  {journal} {Phys. Rev. B}\ }\textbf {\bibinfo {volume} {62}},\ \bibinfo {pages} {9822} (\bibinfo {year} {2000}{\natexlab{b}})}\BibitemShut {NoStop}%
\bibitem [{\citenamefont {Pimenov}\ \emph {et~al.}(2002)\citenamefont {Pimenov}, \citenamefont {Pronin}, \citenamefont {Loidl}, \citenamefont {Tsukada},\ and\ \citenamefont {Naito}}]{pimenov_prb02}%
  \BibitemOpen
  \bibfield  {author} {\bibinfo {author} {\bibfnamefont {A.}~\bibnamefont {Pimenov}}, \bibinfo {author} {\bibfnamefont {A.~V.}\ \bibnamefont {Pronin}}, \bibinfo {author} {\bibfnamefont {A.}~\bibnamefont {Loidl}}, \bibinfo {author} {\bibfnamefont {A.}~\bibnamefont {Tsukada}}, \ and\ \bibinfo {author} {\bibfnamefont {M.}~\bibnamefont {Naito}},\ }\href {\doibase 10.1103/PhysRevB.66.212508} {\bibfield  {journal} {\bibinfo  {journal} {Phys. Rev. B}\ }\textbf {\bibinfo {volume} {66}},\ \bibinfo {pages} {212508} (\bibinfo {year} {2002})}\BibitemShut {NoStop}%
\bibitem [{\citenamefont {Tagay}\ \emph {et~al.}(2021)\citenamefont {Tagay}, \citenamefont {Mahmood}, \citenamefont {Legros}, \citenamefont {Sarkar}, \citenamefont {Greene},\ and\ \citenamefont {Armitage}}]{armitage_prb21}%
  \BibitemOpen
  \bibfield  {author} {\bibinfo {author} {\bibfnamefont {Z.}~\bibnamefont {Tagay}}, \bibinfo {author} {\bibfnamefont {F.}~\bibnamefont {Mahmood}}, \bibinfo {author} {\bibfnamefont {A.}~\bibnamefont {Legros}}, \bibinfo {author} {\bibfnamefont {T.}~\bibnamefont {Sarkar}}, \bibinfo {author} {\bibfnamefont {R.~L.}\ \bibnamefont {Greene}}, \ and\ \bibinfo {author} {\bibfnamefont {N.~P.}\ \bibnamefont {Armitage}},\ }\href {\doibase 10.1103/PhysRevB.104.064501} {\bibfield  {journal} {\bibinfo  {journal} {Phys. Rev. B}\ }\textbf {\bibinfo {volume} {104}},\ \bibinfo {pages} {064501} (\bibinfo {year} {2021})}\BibitemShut {NoStop}%
\bibitem [{\citenamefont {van~der Marel}\ \emph {et~al.}(1993)\citenamefont {van~der Marel}, \citenamefont {Kim}, \citenamefont {Feenstra},\ and\ \citenamefont {Wittlin}}]{vandermarel_comment93}%
  \BibitemOpen
  \bibfield  {author} {\bibinfo {author} {\bibfnamefont {D.}~\bibnamefont {van~der Marel}}, \bibinfo {author} {\bibfnamefont {J.~H.}\ \bibnamefont {Kim}}, \bibinfo {author} {\bibfnamefont {B.~J.}\ \bibnamefont {Feenstra}}, \ and\ \bibinfo {author} {\bibfnamefont {A.}~\bibnamefont {Wittlin}},\ }\href {\doibase 10.1103/PhysRevLett.71.2676} {\bibfield  {journal} {\bibinfo  {journal} {Phys. Rev. Lett.}\ }\textbf {\bibinfo {volume} {71}},\ \bibinfo {pages} {2676} (\bibinfo {year} {1993})}\BibitemShut {NoStop}%
\bibitem [{\citenamefont {Shibauchi}\ \emph {et~al.}(1994)\citenamefont {Shibauchi}, \citenamefont {Kitano}, \citenamefont {Uchinokura}, \citenamefont {Maeda}, \citenamefont {Kimura},\ and\ \citenamefont {Kishio}}]{shibauchi_prl94}%
  \BibitemOpen
  \bibfield  {author} {\bibinfo {author} {\bibfnamefont {T.}~\bibnamefont {Shibauchi}}, \bibinfo {author} {\bibfnamefont {H.}~\bibnamefont {Kitano}}, \bibinfo {author} {\bibfnamefont {K.}~\bibnamefont {Uchinokura}}, \bibinfo {author} {\bibfnamefont {A.}~\bibnamefont {Maeda}}, \bibinfo {author} {\bibfnamefont {T.}~\bibnamefont {Kimura}}, \ and\ \bibinfo {author} {\bibfnamefont {K.}~\bibnamefont {Kishio}},\ }\href {\doibase 10.1103/PhysRevLett.72.2263} {\bibfield  {journal} {\bibinfo  {journal} {Phys. Rev. Lett.}\ }\textbf {\bibinfo {volume} {72}},\ \bibinfo {pages} {2263} (\bibinfo {year} {1994})}\BibitemShut {NoStop}%
\bibitem [{\citenamefont {Panagopoulos}\ \emph {et~al.}(1996)\citenamefont {Panagopoulos}, \citenamefont {Cooper}, \citenamefont {Peacock}, \citenamefont {Gameson}, \citenamefont {Edwards}, \citenamefont {Schmidbauer},\ and\ \citenamefont {Hodby}}]{panagopoulos_prb96}%
  \BibitemOpen
  \bibfield  {author} {\bibinfo {author} {\bibfnamefont {C.}~\bibnamefont {Panagopoulos}}, \bibinfo {author} {\bibfnamefont {J.~R.}\ \bibnamefont {Cooper}}, \bibinfo {author} {\bibfnamefont {G.~B.}\ \bibnamefont {Peacock}}, \bibinfo {author} {\bibfnamefont {I.}~\bibnamefont {Gameson}}, \bibinfo {author} {\bibfnamefont {P.~P.}\ \bibnamefont {Edwards}}, \bibinfo {author} {\bibfnamefont {W.}~\bibnamefont {Schmidbauer}}, \ and\ \bibinfo {author} {\bibfnamefont {J.~W.}\ \bibnamefont {Hodby}},\ }\href {\doibase 10.1103/PhysRevB.53.R2999} {\bibfield  {journal} {\bibinfo  {journal} {Phys. Rev. B}\ }\textbf {\bibinfo {volume} {53}},\ \bibinfo {pages} {R2999} (\bibinfo {year} {1996})}\BibitemShut {NoStop}%
\bibitem [{\citenamefont {Hosseini}\ \emph {et~al.}(2004)\citenamefont {Hosseini}, \citenamefont {Broun}, \citenamefont {Sheehy}, \citenamefont {Davis}, \citenamefont {Franz}, \citenamefont {Hardy}, \citenamefont {Liang},\ and\ \citenamefont {Bonn}}]{bonn_prl04}%
  \BibitemOpen
  \bibfield  {author} {\bibinfo {author} {\bibfnamefont {A.}~\bibnamefont {Hosseini}}, \bibinfo {author} {\bibfnamefont {D.~M.}\ \bibnamefont {Broun}}, \bibinfo {author} {\bibfnamefont {D.~E.}\ \bibnamefont {Sheehy}}, \bibinfo {author} {\bibfnamefont {T.~P.}\ \bibnamefont {Davis}}, \bibinfo {author} {\bibfnamefont {M.}~\bibnamefont {Franz}}, \bibinfo {author} {\bibfnamefont {W.~N.}\ \bibnamefont {Hardy}}, \bibinfo {author} {\bibfnamefont {R.}~\bibnamefont {Liang}}, \ and\ \bibinfo {author} {\bibfnamefont {D.~A.}\ \bibnamefont {Bonn}},\ }\href {\doibase 10.1103/PhysRevLett.93.107003} {\bibfield  {journal} {\bibinfo  {journal} {Phys. Rev. Lett.}\ }\textbf {\bibinfo {volume} {93}},\ \bibinfo {pages} {107003} (\bibinfo {year} {2004})}\BibitemShut {NoStop}%
\bibitem [{\citenamefont {Fazio}\ and\ \citenamefont {{van der Zant}}(2001)}]{fazio_review01}%
  \BibitemOpen
  \bibfield  {author} {\bibinfo {author} {\bibfnamefont {R.}~\bibnamefont {Fazio}}\ and\ \bibinfo {author} {\bibfnamefont {H.}~\bibnamefont {{van der Zant}}},\ }\href {\doibase https://doi.org/10.1016/S0370-1573(01)00022-9} {\bibfield  {journal} {\bibinfo  {journal} {Physics Reports}\ }\textbf {\bibinfo {volume} {355}},\ \bibinfo {pages} {235} (\bibinfo {year} {2001})}\BibitemShut {NoStop}%
\bibitem [{\citenamefont {Paramekanti}\ \emph {et~al.}(2000)\citenamefont {Paramekanti}, \citenamefont {Randeria}, \citenamefont {Ramakrishnan},\ and\ \citenamefont {Mandal}}]{randeria_prb00}%
  \BibitemOpen
  \bibfield  {author} {\bibinfo {author} {\bibfnamefont {A.}~\bibnamefont {Paramekanti}}, \bibinfo {author} {\bibfnamefont {M.}~\bibnamefont {Randeria}}, \bibinfo {author} {\bibfnamefont {T.~V.}\ \bibnamefont {Ramakrishnan}}, \ and\ \bibinfo {author} {\bibfnamefont {S.~S.}\ \bibnamefont {Mandal}},\ }\href {\doibase 10.1103/PhysRevB.62.6786} {\bibfield  {journal} {\bibinfo  {journal} {Phys. Rev. B}\ }\textbf {\bibinfo {volume} {62}},\ \bibinfo {pages} {6786} (\bibinfo {year} {2000})}\BibitemShut {NoStop}%
\bibitem [{\citenamefont {Benfatto}\ \emph {et~al.}(2001)\citenamefont {Benfatto}, \citenamefont {Caprara}, \citenamefont {Castellani}, \citenamefont {Paramekanti},\ and\ \citenamefont {Randeria}}]{benfatto_prb01}%
  \BibitemOpen
  \bibfield  {author} {\bibinfo {author} {\bibfnamefont {L.}~\bibnamefont {Benfatto}}, \bibinfo {author} {\bibfnamefont {S.}~\bibnamefont {Caprara}}, \bibinfo {author} {\bibfnamefont {C.}~\bibnamefont {Castellani}}, \bibinfo {author} {\bibfnamefont {A.}~\bibnamefont {Paramekanti}}, \ and\ \bibinfo {author} {\bibfnamefont {M.}~\bibnamefont {Randeria}},\ }\href {\doibase 10.1103/PhysRevB.63.174513} {\bibfield  {journal} {\bibinfo  {journal} {Phys. Rev. B}\ }\textbf {\bibinfo {volume} {63}},\ \bibinfo {pages} {174513} (\bibinfo {year} {2001})}\BibitemShut {NoStop}%
\bibitem [{\citenamefont {Sun}\ \emph {et~al.}(2020)\citenamefont {Sun}, \citenamefont {Fogler}, \citenamefont {Basov},\ and\ \citenamefont {Millis}}]{millis_prr20}%
  \BibitemOpen
  \bibfield  {author} {\bibinfo {author} {\bibfnamefont {Z.}~\bibnamefont {Sun}}, \bibinfo {author} {\bibfnamefont {M.~M.}\ \bibnamefont {Fogler}}, \bibinfo {author} {\bibfnamefont {D.~N.}\ \bibnamefont {Basov}}, \ and\ \bibinfo {author} {\bibfnamefont {A.~J.}\ \bibnamefont {Millis}},\ }\href {\doibase 10.1103/PhysRevResearch.2.023413} {\bibfield  {journal} {\bibinfo  {journal} {Phys. Rev. Research}\ }\textbf {\bibinfo {volume} {2}},\ \bibinfo {pages} {023413} (\bibinfo {year} {2020})}\BibitemShut {NoStop}%
\bibitem [{\citenamefont {Pimenov}\ \emph {et~al.}(2001)\citenamefont {Pimenov}, \citenamefont {Loidl}, \citenamefont {Duli\ifmmode~\acute{c}\else \'{c}\fi{}}, \citenamefont {van~der Marel}, \citenamefont {Sutjahja},\ and\ \citenamefont {Menovsky}}]{pimenov_prl01}%
  \BibitemOpen
  \bibfield  {author} {\bibinfo {author} {\bibfnamefont {A.}~\bibnamefont {Pimenov}}, \bibinfo {author} {\bibfnamefont {A.}~\bibnamefont {Loidl}}, \bibinfo {author} {\bibfnamefont {D.}~\bibnamefont {Duli\ifmmode~\acute{c}\else \'{c}\fi{}}}, \bibinfo {author} {\bibfnamefont {D.}~\bibnamefont {van~der Marel}}, \bibinfo {author} {\bibfnamefont {I.~M.}\ \bibnamefont {Sutjahja}}, \ and\ \bibinfo {author} {\bibfnamefont {A.~A.}\ \bibnamefont {Menovsky}},\ }\href {\doibase 10.1103/PhysRevLett.87.177003} {\bibfield  {journal} {\bibinfo  {journal} {Phys. Rev. Lett.}\ }\textbf {\bibinfo {volume} {87}},\ \bibinfo {pages} {177003} (\bibinfo {year} {2001})}\BibitemShut {NoStop}%
\bibitem [{\citenamefont {Duli\ifmmode~\acute{c}\else \'{c}\fi{}}\ \emph {et~al.}(2001)\citenamefont {Duli\ifmmode~\acute{c}\else \'{c}\fi{}}, \citenamefont {Pimenov}, \citenamefont {van~der Marel}, \citenamefont {Broun}, \citenamefont {Kamal}, \citenamefont {Hardy}, \citenamefont {Tsvetkov}, \citenamefont {Sutjaha}, \citenamefont {Liang}, \citenamefont {Menovsky}, \citenamefont {Loidl},\ and\ \citenamefont {Saxena}}]{pimenov_vdm_prl01}%
  \BibitemOpen
  \bibfield  {author} {\bibinfo {author} {\bibfnamefont {D.}~\bibnamefont {Duli\ifmmode~\acute{c}\else \'{c}\fi{}}}, \bibinfo {author} {\bibfnamefont {A.}~\bibnamefont {Pimenov}}, \bibinfo {author} {\bibfnamefont {D.}~\bibnamefont {van~der Marel}}, \bibinfo {author} {\bibfnamefont {D.~M.}\ \bibnamefont {Broun}}, \bibinfo {author} {\bibfnamefont {S.}~\bibnamefont {Kamal}}, \bibinfo {author} {\bibfnamefont {W.~N.}\ \bibnamefont {Hardy}}, \bibinfo {author} {\bibfnamefont {A.~A.}\ \bibnamefont {Tsvetkov}}, \bibinfo {author} {\bibfnamefont {I.~M.}\ \bibnamefont {Sutjaha}}, \bibinfo {author} {\bibfnamefont {R.}~\bibnamefont {Liang}}, \bibinfo {author} {\bibfnamefont {A.~A.}\ \bibnamefont {Menovsky}}, \bibinfo {author} {\bibfnamefont {A.}~\bibnamefont {Loidl}}, \ and\ \bibinfo {author} {\bibfnamefont {S.~S.}\ \bibnamefont {Saxena}},\ }\href {\doibase 10.1103/PhysRevLett.86.4144} {\bibfield  {journal} {\bibinfo  {journal} {Phys. Rev. Lett.}\ }\textbf {\bibinfo {volume} {86}},\ \bibinfo {pages} {4144} (\bibinfo {year}
  {2001})}\BibitemShut {NoStop}%
\bibitem [{\citenamefont {Benfatto}\ \emph {et~al.}(2004)\citenamefont {Benfatto}, \citenamefont {Toschi},\ and\ \citenamefont {Caprara}}]{benfatto_prb04}%
  \BibitemOpen
  \bibfield  {author} {\bibinfo {author} {\bibfnamefont {L.}~\bibnamefont {Benfatto}}, \bibinfo {author} {\bibfnamefont {A.}~\bibnamefont {Toschi}}, \ and\ \bibinfo {author} {\bibfnamefont {S.}~\bibnamefont {Caprara}},\ }\href {\doibase 10.1103/PhysRevB.69.184510} {\bibfield  {journal} {\bibinfo  {journal} {Phys. Rev. B}\ }\textbf {\bibinfo {volume} {69}},\ \bibinfo {pages} {184510} (\bibinfo {year} {2004})}\BibitemShut {NoStop}%
\bibitem [{\citenamefont {Fertig}\ and\ \citenamefont {Das~Sarma}(1990)}]{dassarma_prl90}%
  \BibitemOpen
  \bibfield  {author} {\bibinfo {author} {\bibfnamefont {H.~A.}\ \bibnamefont {Fertig}}\ and\ \bibinfo {author} {\bibfnamefont {S.}~\bibnamefont {Das~Sarma}},\ }\href {\doibase 10.1103/PhysRevLett.65.1482} {\bibfield  {journal} {\bibinfo  {journal} {Phys. Rev. Lett.}\ }\textbf {\bibinfo {volume} {65}},\ \bibinfo {pages} {1482} (\bibinfo {year} {1990})}\BibitemShut {NoStop}%
\bibitem [{\citenamefont {Fertig}\ and\ \citenamefont {Das~Sarma}(1991)}]{dassarma_prb91}%
  \BibitemOpen
  \bibfield  {author} {\bibinfo {author} {\bibfnamefont {H.~A.}\ \bibnamefont {Fertig}}\ and\ \bibinfo {author} {\bibfnamefont {S.}~\bibnamefont {Das~Sarma}},\ }\href {\doibase 10.1103/PhysRevB.44.4480} {\bibfield  {journal} {\bibinfo  {journal} {Phys. Rev. B}\ }\textbf {\bibinfo {volume} {44}},\ \bibinfo {pages} {4480} (\bibinfo {year} {1991})}\BibitemShut {NoStop}%
\bibitem [{\citenamefont {Hwang}\ and\ \citenamefont {Das~Sarma}(1995)}]{dassarma_prb95}%
  \BibitemOpen
  \bibfield  {author} {\bibinfo {author} {\bibfnamefont {E.~H.}\ \bibnamefont {Hwang}}\ and\ \bibinfo {author} {\bibfnamefont {S.}~\bibnamefont {Das~Sarma}},\ }\href {\doibase 10.1103/PhysRevB.52.R7010} {\bibfield  {journal} {\bibinfo  {journal} {Phys. Rev. B}\ }\textbf {\bibinfo {volume} {52}},\ \bibinfo {pages} {R7010} (\bibinfo {year} {1995})}\BibitemShut {NoStop}%
\bibitem [{\citenamefont {Melnyk}\ and\ \citenamefont {Harrison}(1970)}]{melnyk_prb70}%
  \BibitemOpen
  \bibfield  {author} {\bibinfo {author} {\bibfnamefont {A.~R.}\ \bibnamefont {Melnyk}}\ and\ \bibinfo {author} {\bibfnamefont {M.~J.}\ \bibnamefont {Harrison}},\ }\href {\doibase 10.1103/PhysRevB.2.835} {\bibfield  {journal} {\bibinfo  {journal} {Phys. Rev. B}\ }\textbf {\bibinfo {volume} {2}},\ \bibinfo {pages} {835} (\bibinfo {year} {1970})}\BibitemShut {NoStop}%
\bibitem [{\citenamefont {Fan}\ \emph {et~al.}(2021)\citenamefont {Fan}, \citenamefont {Yu}, \citenamefont {Cheng}, \citenamefont {Wang}, \citenamefont {Wang}, \citenamefont {Ma}, \citenamefont {Hu}, \citenamefont {Zhang}, \citenamefont {Ma}, \citenamefont {Xue},\ and\ \citenamefont {Song}}]{song_nsr21}%
  \BibitemOpen
  \bibfield  {author} {\bibinfo {author} {\bibfnamefont {J.-Q.}\ \bibnamefont {Fan}}, \bibinfo {author} {\bibfnamefont {X.-Q.}\ \bibnamefont {Yu}}, \bibinfo {author} {\bibfnamefont {F.-J.}\ \bibnamefont {Cheng}}, \bibinfo {author} {\bibfnamefont {H.}~\bibnamefont {Wang}}, \bibinfo {author} {\bibfnamefont {R.}~\bibnamefont {Wang}}, \bibinfo {author} {\bibfnamefont {X.}~\bibnamefont {Ma}}, \bibinfo {author} {\bibfnamefont {X.-P.}\ \bibnamefont {Hu}}, \bibinfo {author} {\bibfnamefont {D.}~\bibnamefont {Zhang}}, \bibinfo {author} {\bibfnamefont {X.-C.}\ \bibnamefont {Ma}}, \bibinfo {author} {\bibfnamefont {Q.-K.}\ \bibnamefont {Xue}}, \ and\ \bibinfo {author} {\bibfnamefont {C.-L.}\ \bibnamefont {Song}},\ }\href {\doibase 10.1093/nsr/nwab225} {\bibfield  {journal} {\bibinfo  {journal} {National Science Review}\ }\textbf {\bibinfo {volume} {9}},\ \bibinfo {pages} {nwab225} (\bibinfo {year} {2021})}\BibitemShut {NoStop}%
\bibitem [{\citenamefont {Mosteller}\ and\ \citenamefont {Wooten}(1968)}]{wooten_fresnel68}%
  \BibitemOpen
  \bibfield  {author} {\bibinfo {author} {\bibfnamefont {L.~P.}\ \bibnamefont {Mosteller}}\ and\ \bibinfo {author} {\bibfnamefont {F.}~\bibnamefont {Wooten}},\ }\href {\doibase 10.1364/JOSA.58.000511} {\bibfield  {journal} {\bibinfo  {journal} {J. Opt. Soc. Am.}\ }\textbf {\bibinfo {volume} {58}},\ \bibinfo {pages} {511} (\bibinfo {year} {1968})}\BibitemShut {NoStop}%
\bibitem [{\citenamefont {Sch\"utzmann}\ \emph {et~al.}(1997)\citenamefont {Sch\"utzmann}, \citenamefont {Somal}, \citenamefont {Tsvetkov}, \citenamefont {van~der Marel}, \citenamefont {Koops}, \citenamefont {Koleshnikov}, \citenamefont {Ren}, \citenamefont {Wang}, \citenamefont {Br\"uck},\ and\ \citenamefont {Menovsky}}]{menovsky_prb97}%
  \BibitemOpen
  \bibfield  {author} {\bibinfo {author} {\bibfnamefont {J.}~\bibnamefont {Sch\"utzmann}}, \bibinfo {author} {\bibfnamefont {H.~S.}\ \bibnamefont {Somal}}, \bibinfo {author} {\bibfnamefont {A.~A.}\ \bibnamefont {Tsvetkov}}, \bibinfo {author} {\bibfnamefont {D.}~\bibnamefont {van~der Marel}}, \bibinfo {author} {\bibfnamefont {G.~E.~J.}\ \bibnamefont {Koops}}, \bibinfo {author} {\bibfnamefont {N.}~\bibnamefont {Koleshnikov}}, \bibinfo {author} {\bibfnamefont {Z.~F.}\ \bibnamefont {Ren}}, \bibinfo {author} {\bibfnamefont {J.~H.}\ \bibnamefont {Wang}}, \bibinfo {author} {\bibfnamefont {E.}~\bibnamefont {Br\"uck}}, \ and\ \bibinfo {author} {\bibfnamefont {A.~A.}\ \bibnamefont {Menovsky}},\ }\href {\doibase 10.1103/PhysRevB.55.11118} {\bibfield  {journal} {\bibinfo  {journal} {Phys. Rev. B}\ }\textbf {\bibinfo {volume} {55}},\ \bibinfo {pages} {11118} (\bibinfo {year} {1997})}\BibitemShut {NoStop}%
\end{thebibliography}%

\end{document}